\documentclass{aastex63}

\usepackage{ulem}

\newcommand\subs[1]{\textsubscript{#1}}
\newcommand\sups[1]{\textsuperscript{#1}}

\definecolor{gold}{rgb}{0.64,0.54,0.29}

\received{2020 October 1}
\revised{2020 December 18}
\accepted{2021 January 14}
\submitjournal{PSJ}

\shorttitle{}
\shortauthors{Roth et al.}

\begin{document}

\title{Rapidly Varying Anisotropic Methanol (CH$_3$OH) Production in the Inner Coma of Comet 46P/Wirtanen as Revealed by the ALMA Atacama Compact Array}

\correspondingauthor{Nathan X. Roth}
\email{nathaniel.x.roth@nasa.gov}

\author[0000-0002-6006-9574]{Nathan X. Roth}
\altaffiliation{Visiting Astronomer at the Infrared Telescope Facility, which is operated by the University of Hawaii under contract NNH14CK55B with the National Aeronautics and Space Administration.}
\affiliation{Solar System Exploration Division, Astrochemistry Laboratory Code 691, NASA Goddard Space Flight Center, 8800 Greenbelt Rd, Greenbelt, MD 20771, USA}
\affiliation{Universities Space Research Association, Columbia, MD 21064, USA}

\author [0000-0001-7694-4129]{Stefanie N. Milam}
\affiliation{Solar System Exploration Division, Astrochemistry Laboratory Code 691, NASA Goddard Space Flight Center, 8800 Greenbelt Rd, Greenbelt, MD 20771, USA}

\author[0000-0001-8233-2436]{Martin A. Cordiner}
\affiliation{Solar System Exploration Division, Astrochemistry Laboratory Code 691, NASA Goddard Space Flight Center, 8800 Greenbelt Rd, Greenbelt, MD 20771, USA}
\affiliation{Department of Physics, Catholic University of America, Washington DC, USA}

\author{Dominique Bockelée-Morvan}
\affiliation{LESIA, Observatoire de Paris, Université PSL, CNRS, Sorbonne Université,
Université de Paris, 5 place Jules Janssen, F-92195 Meudon, France}

\author[0000-0001-8843-7511]{Michael A. DiSanti}
\altaffiliation{Visiting Astronomer at the Infrared Telescope Facility, which is operated by the University of Hawaii under contract NNH14CK55B with the National Aeronautics and Space Administration.}
\affiliation{Solar System Exploration Division, Planetary Systems Laboratory Code 693, NASA Goddard Space Flight Center, 8800 Greenbelt Rd, Greenbelt, MD 20771, USA}

\author[0000-0002-1545-2136]{Jérémie Boissier}
\affiliation{Institut de Radioastronomie Millimetrique, 300 rue de la piscine, F-38406
Saint Martin d'Heres, France}

\author[0000-0003-2414-5370]{Nicolas Biver}
\affiliation{LESIA, Observatoire de Paris, Université PSL, CNRS, Sorbonne Université,
Université de Paris, 5 place Jules Janssen, F-92195 Meudon, France}

\author{Jacques Crovisier}
\affiliation{LESIA, Observatoire de Paris, Université PSL, CNRS, Sorbonne Université,
Université de Paris, 5 place Jules Janssen, F-92195 Meudon, France}

\author[0000-0002-8379-7304]{Neil Dello Russo}
\altaffiliation{Visiting Astronomer at the Infrared Telescope Facility, which is operated by the University of Hawaii under contract NNH14CK55B with the National Aeronautics and Space Administration.}
\affiliation{Johns Hopkins University Applied Physics Laboratory, 11100 Johns Hopkins Rd, Laurel, MD 20723, USA}

\author[0000-0002-6391-4817]{Boncho P. Bonev}
\altaffiliation{Visiting Astronomer at the Infrared Telescope Facility, which is operated by the University of Hawaii under contract NNH14CK55B with the National Aeronautics and Space Administration.}
\affiliation{Department of Physics, American University, 4400 Massachusetts Avenue NW, Washington, DC, 20016 USA}

\author[0000-0001-8642-1786]{Chunhua Qi}
\affiliation{Harvard-Smithsonian Center for Astrophysics, 60 Garden Street, Mail Stop 42, Cambridge, MA 02138, USA}

\author[0000-0001-9479-9287]{Anthony J. Remijan}
\affiliation{National Radio Astronomy Observatory, 520 Edgemont Rd, Charlottesville, VA 22903, USA}

\author[0000-0001-6752-5109]{Steven B. Charnley}
\affiliation{Solar System Exploration Division, Astrochemistry Laboratory Code 691, NASA Goddard Space Flight Center, 8800 Greenbelt Rd, Greenbelt, MD 20771, USA}

\author[0000-0003-0142-5265]{Erika L. Gibb}
\altaffiliation{Visiting Astronomer at the Infrared Telescope Facility, which is operated by the University of Hawaii under contract NNH14CK55B with the National Aeronautics and Space Administration.}
\affiliation{Department of Physics \& Astronomy, 1 University Blvd, University of Missouri-St.Louis, St. Louis, MO, 63121 USA}

\author[0000-0002-0455-9384]{Miguel de Val-Borro}
\affiliation{Planetary Science Institute, 1700 E. Fort Lowell, Suite 106, Tucson, AZ 85719, USA}

\author[0000-0001-8923-488X]{Emmanu\"{e}l Jehin}
\affiliation{Space sciences, Technologies \& Astrophysics Research (STAR) Institute, University of Liège, Belgium}



\begin{abstract}
We report the detection of CH$_3$OH emission in  comet 46P/Wirtanen on UT 2018 December 8 and 9 using the  Atacama Compact Array (ACA), part of the Atacama Large Millimeter/Submillimeter Array (ALMA). These interferometric measurements of CH$_3$OH along with continuum emission from  dust probed the inner coma ($<$2000 km from the nucleus) of 46P/Wirtanen approximately one week before its closest approach to Earth ($\Delta$ = 0.089 -- 0.092 au), revealing rapidly varying and anisotropic CH$_3$OH outgassing during five separate ACA executions between UT 23:57 December 7 and UT 04:55 December 9, with a clear progression in the spectral line profiles over a timescale of minutes. We present spectrally integrated flux maps, production rates, rotational temperatures, and spectral line profiles of CH$_3$OH during each ACA execution. The variations in CH$_3$OH outgassing are consistent with Wirtanen's 9 hr nucleus rotational period derived from optical and millimeter wavelength measurements and thus are likely coupled to the changing illumination of active sites on the nucleus. The consistent blue offset of the line center indicates enhanced CH$_3$OH sublimation from the sunward hemisphere of the comet, perhaps from icy grains.  These results demonstrate the exceptional capabilities of the ACA for time-resolved measurements of comets such as 46P/Wirtanen.

\end{abstract}



\section{Introduction} \label{sec:intro}

Comets, kilometer-sized bodies of ice and dust, were assembled in the solar nebula during the era of planet formation and subsequently scattered to the Kuiper disk or the Oort cloud. Stored with minimal processing for the last $\sim$4.5 Gyr in the deep freeze of the cold outer solar system, they may serve as ``fossils'' of solar system formation, with the volatile composition of their nuclei reflecting the chemistry and prevailing conditions present in the midplane of the solar nebula \citep{Bockelee2004,Mumma2011a,Bockelee2017}. Volatiles are released as they enter the inner solar system (heliocentric distances, \textit{r}\subs{H} $<$ 5 au) and encounter increasing solar radiation, activating sublimation and leading to the formation of a coma (a freely expanding atmosphere). Remote sensing of coma volatiles over the past three decades has revealed that comets display significant compositional diversity \citep[e.g.,][]{AHearn1995,Crovisier2009,Cochran2015,DelloRusso2016a}, providing a window into the diversity of their initial conditions and the degree of chemical processing during their lifetime. 

The composition and structure of the coma can be measured in exceptional detail through pure rotational transitions of coma molecules, such as carbon monoxide (CO), carbon monosulfide (CS), hydrogen cyanide (HCN), formaldehyde (H$_2$CO), and methanol (CH$_3$OH). Single-dish and interferometric observations at millimeter wavelengths \citep[e.g.,][]{Bockelee1994,Biver1999,Fray2006,Milam2006,Boissier2007,Lis2008,Cordiner2014} have shown that the distributions of HCN and CH$_3$OH in some comets are consistent with direct release from the nucleus, whereas CS, H$_2$CO, and HNC are produced by ``distributed'' sources in the coma. More recently, the high sensitivity and angular resolution of the Atacama Large Millimeter/Submillimeter Array (ALMA) have confirmed these results, and enabled spatially and temporally resolved measurements capable of discerning variations in outgassing on small timescales \citep{Cordiner2017b}. 

\begin{deluxetable*}{cccccccccccc}[h]
\tablenum{1}
\tablecaption{Observing Log\label{tab:obslog}}
\tablewidth{0pt}
\tablehead{
\colhead{Execution} & \colhead{Date} & \colhead{UT Time} & \colhead{\textit{T}\subs{int}} &
\colhead{\textit{r}\subs{H}} & \colhead{$\Delta$} & \colhead{$\phi$} & \colhead{$\nu$} & 
\colhead{\textit{N}\subs{ants}}  & \colhead{Baselines} & \colhead{$\theta$\subs{min}} & \colhead{PWV} \\
\colhead{} & \colhead{} & \nocolhead{} & \colhead{(min)} & \colhead{(au)} & 
\colhead{(au)} & \colhead{($\degr$)} & \colhead{(GHz)}  & \colhead{} & \colhead{(m)}  &
\colhead{($\arcsec$)} & \colhead{(mm)}
}
\startdata
1 & 2018 Dec 7--8 & 23:57--00:58 & 45 & 1.057 & 0.092 & 37.1 & 241.8 & 12 & 8.9--48.9 & 4.11$\times$7.99 & 1.09 \\
2 & 2018 Dec 8 & 01:26--02:26 & 45 & 1.057 & 0.092 & 37.0 & 241.8 & 12 & 8.8--48.9 & 3.98$\times$6.53 & 0.89 \\
3 & 2018 Dec 9 & 00:46--01:47 & 45 & 1.056 & 0.089 & 35.3 & 241.8 & 12 & 8.8--48.9 & 4.23$\times$7.27 & 2.15 \\
4 & 2018 Dec 9 & 02:26--03:27 & 45 & 1.056 & 0.089 & 35.1 & 241.8 & 12 & 8.8--48.9 & 4.28$\times$6.85 & 2.58 \\
5 & 2018 Dec 9 & 03:54--04:55 & 45 & 1.056 & 0.089 & 35.0 & 241.8 & 12 & 8.8--48.9 & 4.33$\times$7.78 & 3.23 \\
\enddata
\tablecomments{\textit{T}\subs{int} is the total on-source integration time. \textit{r}\subs{H}, $\Delta$, and $\phi$ are the heliocentric distance, geocentric distance,
and phase angle (Sun--Comet--Earth), respectively, of Wirtanen at the time of observations. $\nu$ is the mean frequency of each instrumental setting. \textit{N}\subs{ants}
is the number of antennas utilized during each observation, with the range of baseline lengths indicated for each. $\theta$\subs{min} is the angular resolution (synthesized beam) at $\nu$, and PWV is the mean precipitable water vapor at zenith during the observations.}
\end{deluxetable*}

The 2018 perihelion passage of the Jupiter-family comet (JFC) 46P/Wirtanen afforded an opportunity to couple the high spatial and spectral resolution of ALMA with the most favorable apparition of a JFC in the previous decade. Here we report time-resolved pre-perihelion observations of CH$_3$OH in 46P/Wirtanen taken with the ALMA Atacama Compact Array (ACA). The 12 $\times$ 7 m antennas of the ACA provide short baselines between 9 and $\sim$50 m (corresponding to an angular resolution ranging from 5$\farcs$45 to 29$\farcs$0 at 230 GHz), resulting in greater sensitivity to extended flux than the main 12 m array (which was in configuration C43-4 at the time of our observations with an angular resolution ranging from 0$\farcs$39 to 4$\farcs$89 at 230 GHz). We note that these observations were most sensitive to the inner coma of 46P/Wirtanen, and that the largest recoverable scale of the ACA is small compared to the size of 46P/Wirtanen's extended coma (up to several arcminutes). We sampled CH$_3$OH emission in the coma of 46P/Wirtanen with multiple executions of the ACA spread over two dates, searching for changes in molecular production rates, outgassing, and spatial distribution of the emission. Significant variations in molecular production and anisotropic outgassing of CH$_3$OH were measured throughout these data. We report spectrally integrated flux maps, production rates, rotational temperatures, and spectral line profiles of CH$_3$OH emission. Section~\ref{sec:obs} presents the observations and data analysis. Results are found in Section~\ref{sec:results}. Section~\ref{sec:variations} discusses the variation in CH$_3$OH during these measurements, identifies potential mechanisms, and places these results in the context of other comets characterized to date. 

\begin{figure}[h!]
\gridline{\fig{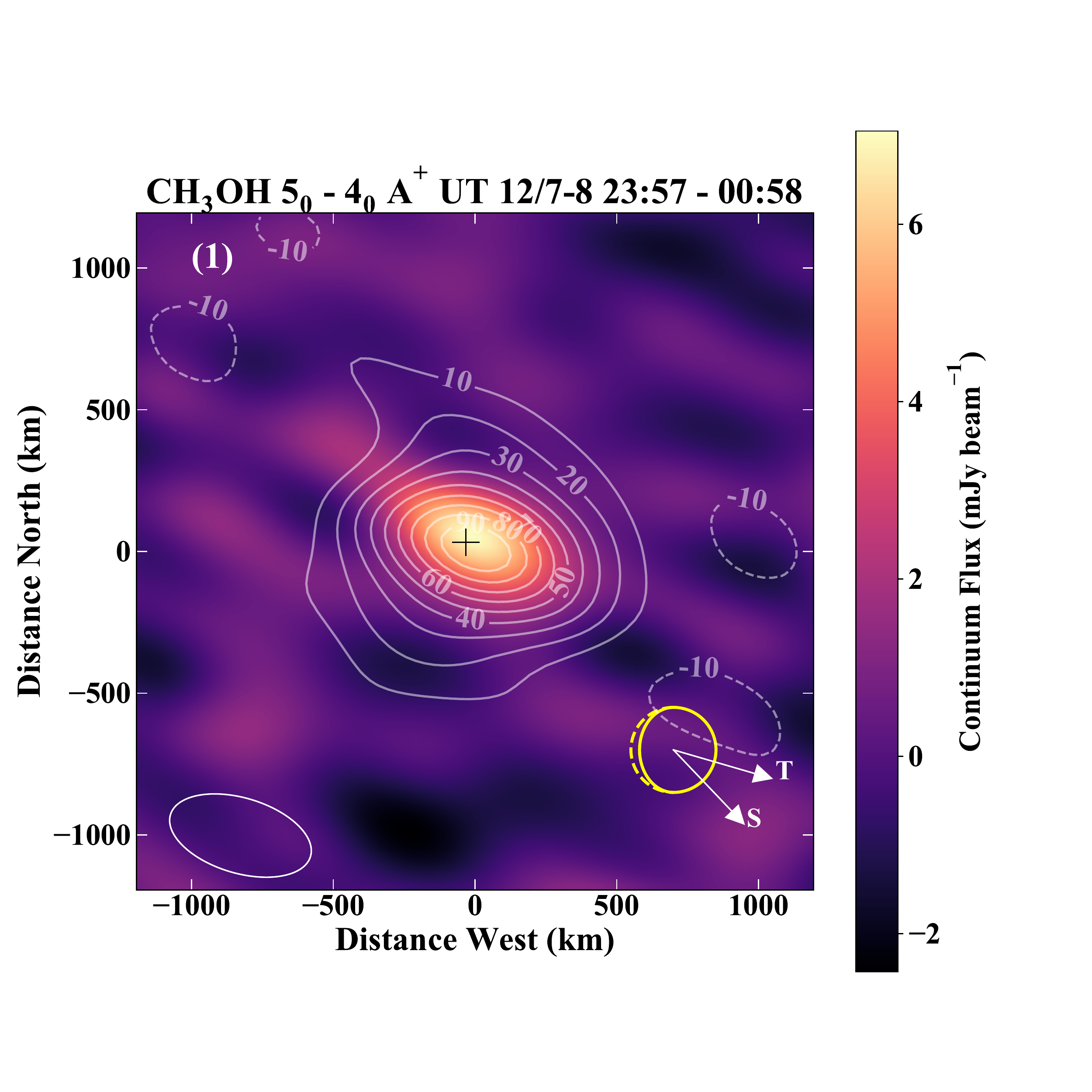}{0.48\textwidth}{(A)}
          \fig{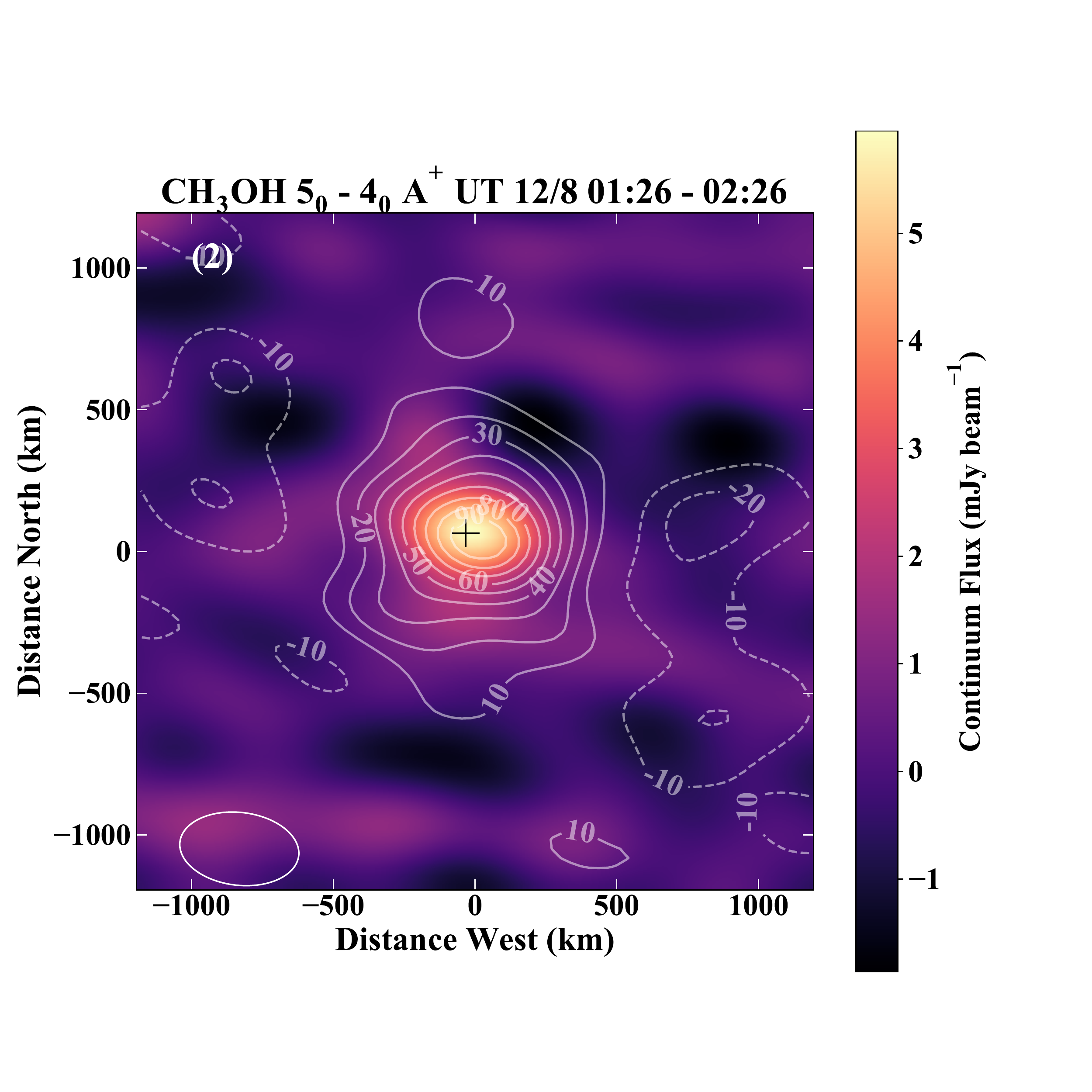}{0.48\textwidth}{(B)}
	}
\caption{Executions 1--2. Spectrally integrated flux maps for the CH$_3$OH $5_0$--$4_0$ $A^{+}$ transition in Wirtanen on UT 2018 December 8. Contour intervals are 10\% of the projected peak of the gas emission in each map. The rms noise ($\sigma$, mJy beam$^{-1}$ km s$^{-1}$) and contour spacing, $\delta$, for each map are (A): $\sigma$ = 19, $\delta$ = 2.7$\sigma$ and (B): $\sigma$ = 18, $\delta$ = 2.2$\sigma$. Continuum emission is shown as a color map behind the contours. Sizes and orientations of the synthesized beam are indicated in the lower-left corner of each panel. The comet's illumination phase ($\phi \sim$ 36$\degr$), as well as the direction of the Sun and dust trail, are indicated in the lower-right. The black cross indicates the position of the peak continuum flux from which spectra were extracted. Each execution is numbered in the upper left as denoted in Table~\ref{tab:obslog}.
\label{fig:maps1}}
\end{figure}

\section{Observations and Data Reduction}\label{sec:obs}
Comet 46P/Wirtanen (hereafter Wirtanen) is a JFC with a period \textit{P} = 5.4 yr, and was the original target of ESA's Rosetta mission. During its 2018 apparition, Wirtanen reached perihelion (\textit{q} = 1.05 au) on UT 2018 December 12 and passed closest to the Earth ($\Delta$\subs{min} = 0.0774 au, approximately 30 lunar distances) on UT 2018 December 16. We conducted pre-perihelion observations toward Wirtanen on UT 2018 December 8 and 9 during Cycle 6 using the ALMA ACA with the Band 6 receiver, covering frequencies between 241.5 and 242.0 GHz ($\lambda$ = 1.23 -- 1.24 mm) in one 500 MHz wide spectral window. The observing log is shown in Table~\ref{tab:obslog}. We tracked the comet position using JPL Horizons ephemerides (JPL \#K181/21). One correlator setting was employed in each execution to simultaneously sample multiple CH$_3$OH transitions and continuum. Weather conditions were good (mean precipitable water vapor at zenith, zenith PWV, = 0.89 -- 1.09 mm on December 8; zenith PWV = 2.15 -- 3.23 mm on December 9). Quasar observations were used for bandpass and phase calibration, as well as calibrating Wirtanen's flux scale. The spatial scale (the range in semi-minor and semi-major axes of the synthesized beam) was 3$\farcs$98 -- 7$\farcs$99 and the channel spacing was 122 kHz, leading to a spectral resolution of 0.15 km s$^{-1}$. The data were flagged, calibrated, and imaged using standard routines in Common Astronomy Software Applications (CASA) package version 5.6.1 \citep{McMullin2007}. We deconvolved the point-spread function with the Högbom algorithm, using natural visibility weighting and a flux threshold of twice the rms noise in each image. The deconvolved images were then convolved with the synthesized beam and corrected for the (Gaussian) response of the ALMA primary beam. We transformed the images from astrometric coordinates to projected cometocentric distances, with the location of the peak continuum flux chosen as the origin, which was in good agreement with the comet's predicted ephemeris position. The topocentric frequency of each spectral channel was converted to cometocentric velocity as
\begin{equation}
    V = c\left(\frac{f_{rest}-f}{f_{rest}}\right) - \frac{d\Delta}{dt}
\end{equation}
where c is the speed of light (km s$^{-1}$), \textit{f}\subs{rest} is the rest frequency of a given transition (GHz), \textit{f} is the frequency of the channel (GHz), and d$\Delta$/dt is the topocentric velocity of the comet at the time of the observations (km s$^{-1}$).

\section{Results}\label{sec:results}
We detected molecular emission from multiple strong CH$_3$OH transitions near 241.7 GHz as well as continuum emission from dust in the coma of Wirtanen. Molecular line emission was modeled using a three-dimensional radiative transfer method based on the Line Modeling Engine \citep[LIME;][]{Brinch2010} adapted for cometary atmospheres, including a full non-LTE treatment of coma gases, collisions with H$_2$O and electrons, and pumping by solar radiation \citep[for further details, see][]{Cordiner2019}. We calculated the number density of molecules released from the nucleus as a function of distance following the Haser formalism \citep{Haser1957} as
\begin{equation}
    n_p(r) = \frac{Q}{4 \pi v_{exp} r^2}\exp{\left(-\frac{r\beta}{v_{exp}}\right)},
\end{equation}
\noindent where \textit{Q} is the production rate (molecules s$^{-1}$), \textit{v}\subs{exp} is the expansion velocity (km s$^{-1}$) and \textit{$\beta$} is the photodissociation rate adopted from \cite{Huebner1992}, with \textit{Q} and \textit{v}\subs{exp} allowed to vary as free parameters.

We assumed that the collision rate between CH$_3$OH and H$_2$O is the same as that with H$_2$ \citep[taken from the LAMDA database;][]{Schoier2005}. This assumption has negligible impact on the derived parameters as the coma contained within the ACA beam was close to LTE due to the proximity of Wirtanen and the relatively small beam size (Section~\ref{subsec:trot}). Our models were run through the CASA ``simobserve'' tool to mimic the effects of the ACA using the exact antenna configuration and hour angle ranges of our observations. We found that interferometric filtering by the ACA resulted in a 24\% loss in the peak flux of our models; therefore, an efficiency factor $\eta$ = 0.76 was applied to the radiative transfer models when comparing with observed spectra and calculating production rates.   Detected spectral lines, including upper-state energies (\textit{E}\subs{u}) and integrated fluxes, are listed in Table~\ref{tab:lines}.

\begin{figure*}[h]
\gridline{\fig{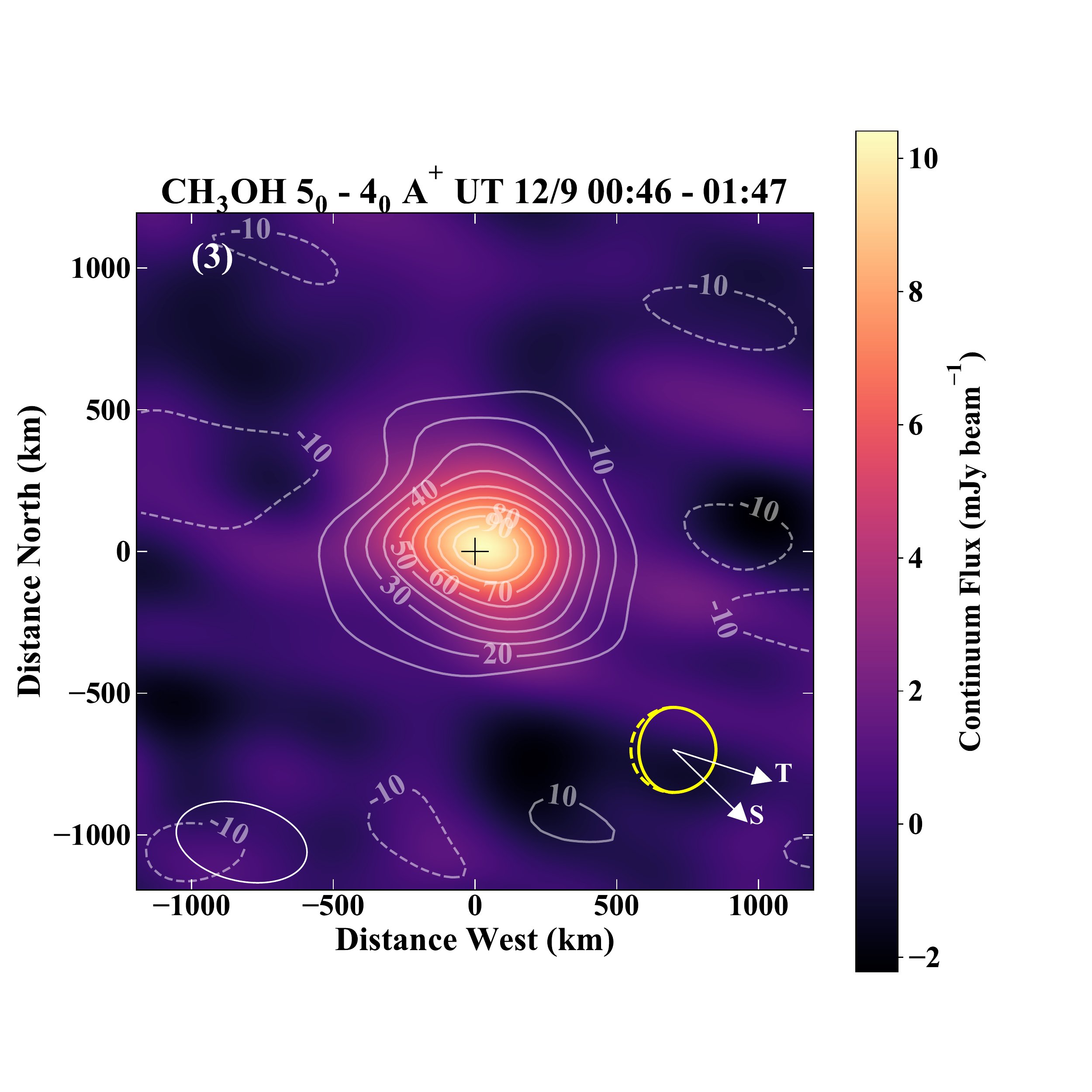}{0.48\textwidth}{(A)}
          \fig{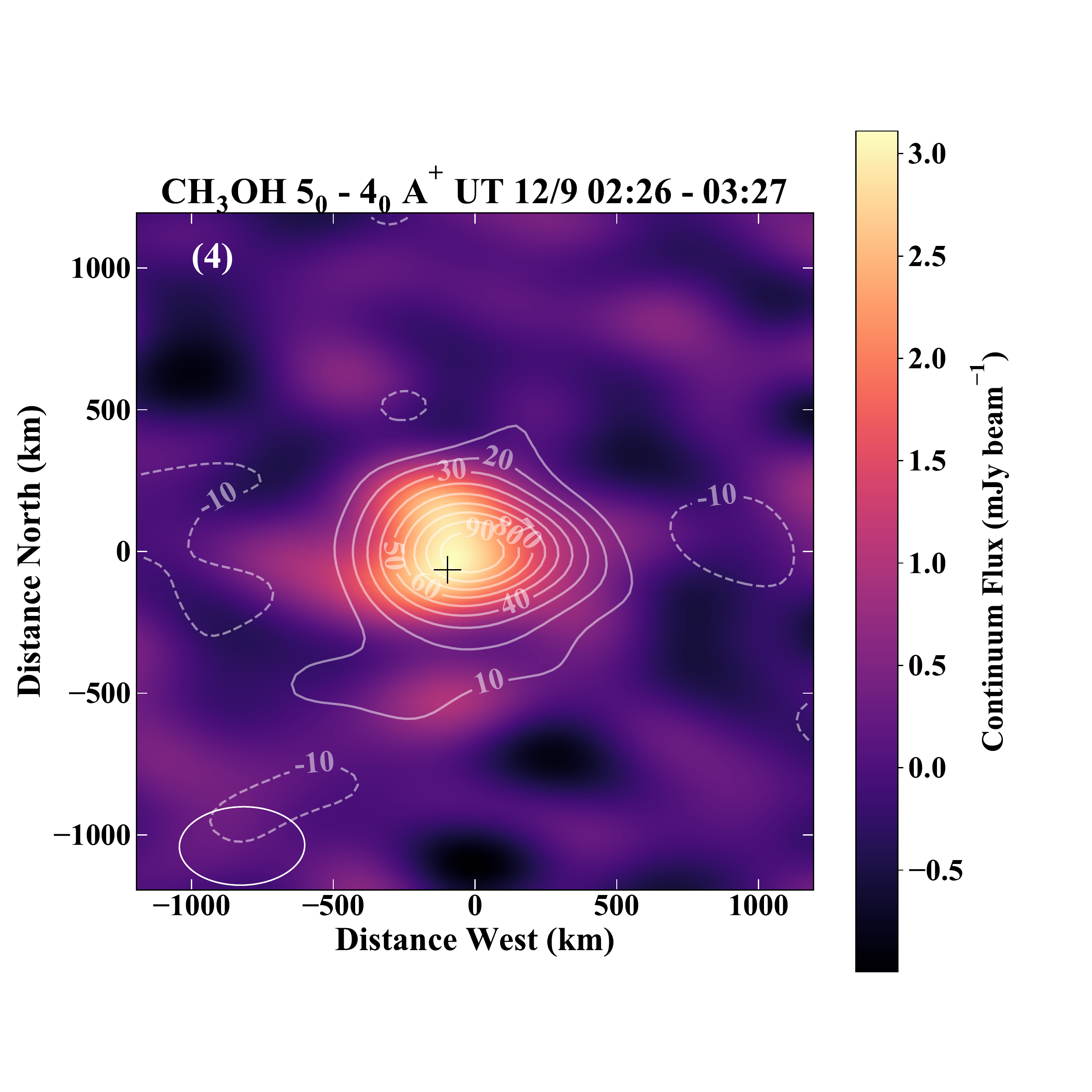}{0.48\textwidth}{(A)}
          }
\gridline{\fig{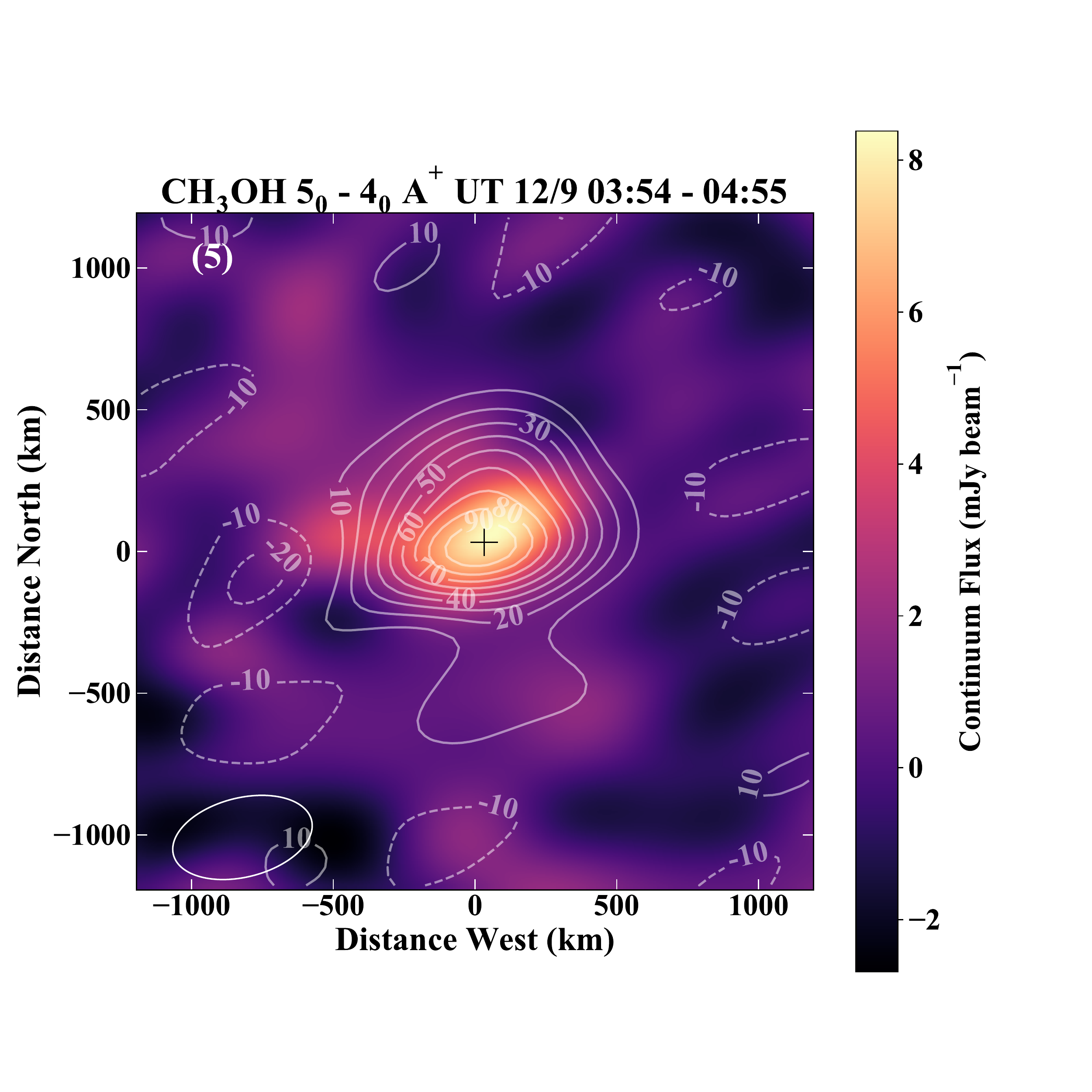}{0.48\textwidth}{(C)}
          }

\caption{Executions 3--5. Spectrally integrated flux maps for the CH$_3$OH $5_0$--$4_0$ $A^{+}$ transition in Wirtanen on UT 2018 December 9, with traces and labels as in Figure~\ref{fig:maps1}. Contour intervals are 10\% of the projected peak of the gas emission in each map. The rms noise ($\sigma$, mJy beam$^{-1}$ km s$^{-1}$) and contour spacing, $\delta$, for each map are (A): $\sigma$ = 16, $\delta$ = 3.3$\sigma$, (B): $\sigma$ = 18, $\delta$ = 3.3$\sigma$, and (C): $\sigma$ = 21, $\delta$ = 2.7$\sigma$. Continuum emission is shown as a color map behind the contours. Each execution is numbered in the upper left as denoted in Table~\ref{tab:obslog}.
\label{fig:maps2}}
\end{figure*}

\begin{deluxetable*}{cccccccc}
\tablenum{2}
\tablecaption{CH\subs{3}OH Lines Detected in Wirtanen\label{tab:lines}}
\tablewidth{0pt}
\tablehead{
\colhead{Transition} & \colhead{Frequency} & \colhead{\textit{E}\subs{u}} & \multicolumn{5}{c}{Integrated Flux\sups{a}} \\
\colhead{} & \colhead{(GHz)} & \colhead{(K)} & \multicolumn{5}{c}{(K km s\sups{-1})}
}
\startdata
 Execution & & & 1 & 2 & 3 & 4 & 5 \\
 UT Date (2018) & & & 12/7-8, 23:57--00:58 & 12/8, 01:26--02:26 & 12/9, 00:46--01:47 & 12/9, 02:26--03:27 & 12/9, 03:54--04:55 \\
\hline
$5_0$--$4_0$ $A^{+}$ & 241.791 & 34.8 & 0.381 $\pm$ 0.011 & 0.305 $\pm$ 0.015 & 0.349 $\pm$ 0.013 & 0.441 $\pm$ 0.012 & 0.345 $\pm$ 0.014 \\
$5_{-1}$--$4_{-1}$ $E$ & 241.767 & 40.4 & 0.317 $\pm$ 0.010 & 0.278 $\pm$ 0.014 & 0.337 $\pm$ 0.015 & 0.360 $\pm$ 0.014 & 0.270 $\pm$ 0.018 \\
$5_0$--$4_0$ $E$ & 241.700 & 47.9 & 0.313 $\pm$ 0.011 & 0.279 $\pm$ 0.013 & 0.293 $\pm$ 0.015 & 0.357 $\pm$ 0.015 & 0.259 $\pm$ 0.014\\
$5_1$--$4_1$ $E$ & 241.879 & 55.9 & 0.260 $\pm$ 0.010 & 0.244 $\pm$ 0.015 & 0.243 $\pm$ 0.012 & 0.283 $\pm$ 0.013 & 0.242 $\pm$ 0.016\\
$5_2$--$4_2$ $A^{+}$ & 241.887 & 72.5 & 0.215 $\pm$ 0.009 & 0.149 $\pm$ 0.011 & 0.122 $\pm$ 0.014 & 0.195 $\pm$ 0.014 & 0.078 $\pm$ 0.019\\
$5_{-3}$--$4_{-3}$ $E$ & 241.852 & 97.5 & 0.080 $\pm$ 0.011 & 0.093 $\pm$ 0.013 & 0.078 $\pm$ 0.013 & 0.108 $\pm$ 0.014 & 0.084 $\pm$ 0.015\\
\textit{T}\subs{rot}\sups{b} (K) & & & 82 $\pm$ 8 & 77 $\pm$ 10 & 52 $\pm$ 5 & 60 $\pm$ 5 & 57 $\pm$ 7 \\
\hline
\enddata
\tablecomments{\sups{a} Spectrally integrated flux of spectra extracted from a nucleus-centered beam at the position of the peak continuum flux (Figures~\ref{fig:maps1} and~\ref{fig:maps2}) for the indicated execution and UT date labeled in Table~\ref{tab:obslog}. \sups{b} Rotational temperature derived using the rotational diagram method \citep{Bockelee1994}.
}
\end{deluxetable*}

\subsection{Molecular Maps and Extracted Spectra}\label{subsec:maps}
Figures~\ref{fig:maps1} and~\ref{fig:maps2} show spectrally integrated flux maps for the CH$_3$OH $5_0$--$4_0$ $A^{+}$ transition near 241.791 GHz (the strongest CH$_3$OH transition in these data) for each ACA execution in Wirtanen. The shape and spatial extent of the emissions are relatively consistent across each execution. 

\begin{figure*}[h]
\plotone{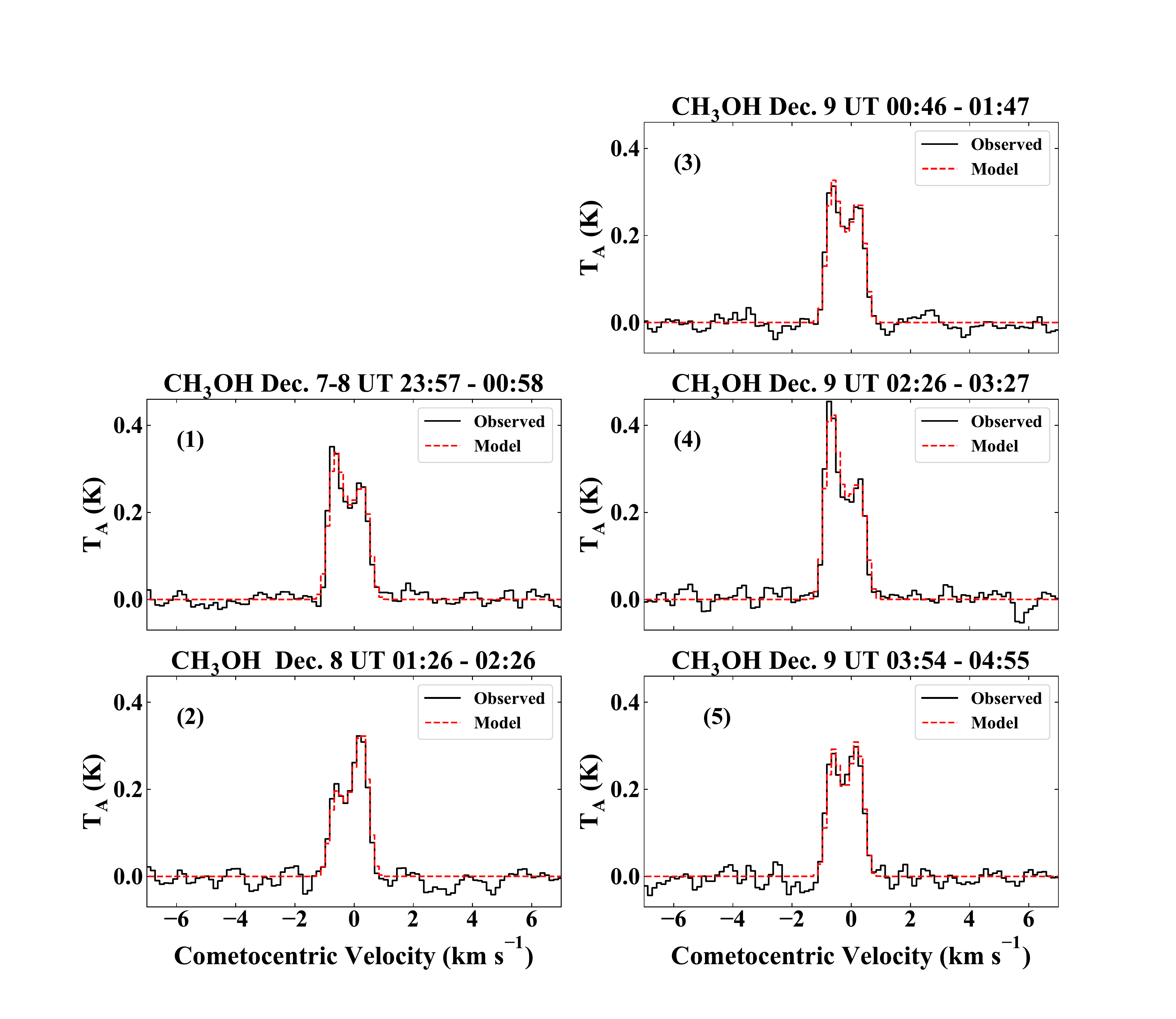}
\caption{Extracted spectra of the CH$_3$OH $5_0$--$4_0$ $A^{+}$ transition in Wirtanen during each ACA execution with best-fit models overlaid. Best-fit model parameters are given in Table~\ref{tab:vexp}. Each execution is numbered as denoted in Table~\ref{tab:obslog}. \label{fig:spectra}}
\end{figure*}

We extracted nucleus-centered spectra of the $5_0$--$4_0$ $A^{+}$ transition in each ACA execution. Figure~\ref{fig:spectra} shows the significant variation in the velocity profile (and to an extent, amplitude) measured in each execution. We found that the velocity profiles of the remaining five detected CH$_3$OH transitions in each execution (Table~\ref{tab:lines}) were consistent with those of the strongest $5_0$--$4_0$ $A^{+}$ transition. Taken together, these spectra indicate that CH$_3$OH outgassing in Wirtanen was anisotropic, and furthermore, that this anisotropy varied with time (Section~\ref{sec:variations}).

\subsection{Rotational and Kinetic Temperatures}\label{subsec:trot}
We detected emission from six strong CH$_3$OH transitions in each ACA execution on December 8 and 9. We constrained the coma rotational temperature along the line of sight during each execution using the rotational diagram method \citep{Bockelee1994}, and list them in Table~\ref{tab:lines}. Retrieved rotational temperatures are marginally higher on December 8 than on December 9. However, rotational temperatures for executions on a given date are consistent within 1$\sigma$ uncertainty. 

Given the very small geocentric distance ($\Delta \sim$ 0.09 au) of Wirtanen during our observations, the half-width at half-maximum of the ACA synthesized central beam extended to nucleocentric distances of $\sim$132 $\times$  265 km. This corresponds to the very inner coma of Wirtanen, where the rotational states should be in thermal equilibrium due to collisions at the kinetic temperature of the gas; therefore, the rotational temperature is expected to be similar to the kinetic temperature for the CH$_3$OH sampled by our measurements. 

\begin{figure*}[h]
\plotone{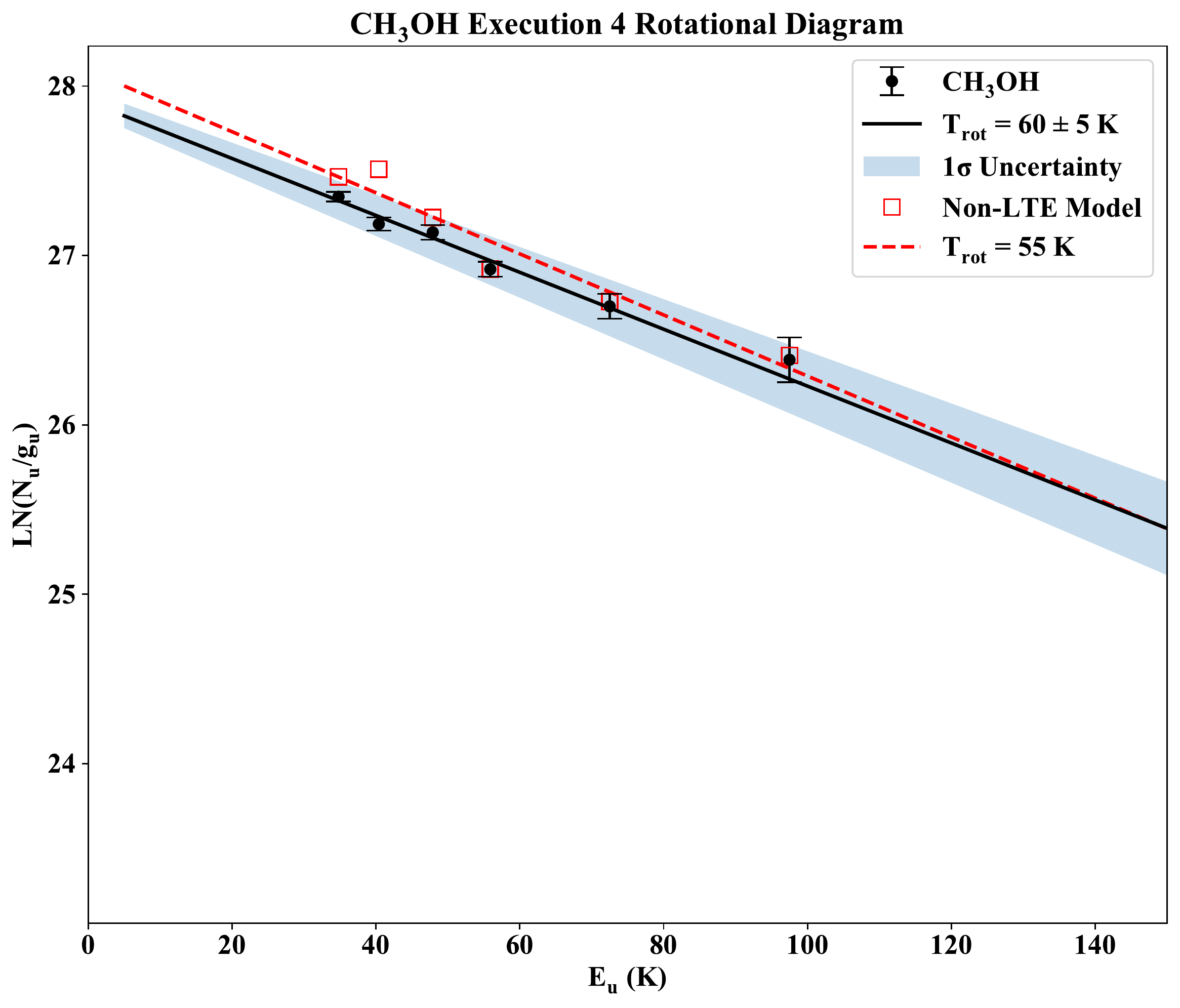}
\caption{Rotational diagram of the CH$_3$OH $5_0$--$4_0$ $A^{+}$, $5_{-1}$--$4_{-1}$ $E$, $5_0$--$4_0$ $E$, $5_1$--$4_1$ $E$, $5_2$--$4_2$ $A^{+}$, and $5_{-3}$--$4_{-3}$ $E$ transitions in Wirtanen on UT 2018 December 9 during Execution 4 (black circles, Tables~\ref{tab:obslog} and~\ref{tab:lines}), along with the best-fit line (black, solid), corresponding to \textit{T}\subs{rot} = 60 $\pm$ 5 K, and the 1$\sigma$ uncertainty in the fit (blue shaded area). Also shown is the rotational diagram for a non-LTE model with \textit{T}\subs{kin} = 60 K (red squares) and the best-fit line (red, dashed) corresponding to \textit{T}\subs{rot} = 55 K, illustrating formal agreement between the observations and the model.}\label{fig:trot}
\end{figure*}

We tested this by running full non-LTE models for all six detected CH$_3$OH transitions in Execution 4 (the execution with the strongest transitions and highest signal-to-noise ratio, S/N), setting the model kinetic temperature equal to the best-fit rotational temperature for the execution (\textit{T}\subs{rot} = 60 $\pm$ 5 K; Table~\ref{tab:lines}). We then performed a rotational diagram analysis on the model-generated lines. We find \textit{T}\subs{rot}(model) = 55 K, in formal agreement with the rotational temperature calculated from the observed spectra (Figure~\ref{fig:trot}). Thus, the use of LTE models (i.e., setting the kinetic temperature of the model to the derived rotational temperature for each execution) is justified and results in negligible changes in production rate (8\% difference for $\pm$5 K changes in \textit{T}). Given that these conditions were satisfied for Execution 4, it is reasonable to assume that they were satisfied for the other executions. These results are in excellent agreement with data obtained with the IRAM 30 m \citep{Biver2021}.

\subsection{Expansion Velocities and Anisotropic Outgassing}
We derived the expansion velocities and outgassing asymmetry factors for the CH$_3$OH $5_0$--$4_0$ $A^{+}$ transition in each execution by performing nonlinear least-squares fits of our radiative transfer models to extracted spectral line profiles. Hemispheric asymmetry in outgassing along the Sun--comet vector was considered when fitting the observations for each execution. Consistent with previous ALMA studies of comets, we modeled CH$_3$OH as a parent species \citep{Cordiner2017a,Cordiner2019b}. These models produced excellent fits to the data (reduced chi-square, $\chi^2$\subs{red} = 0.97--1.35) (see Figure~\ref{fig:spectra}), allowing the spectral line asymmetry to be fully reproduced. Table~\ref{tab:vexp} lists our results, and Figure~\ref{fig:spectra} shows our extracted spectra with best-fit models overlain.

\begin{deluxetable*}{cccccccc}
\tablenum{3}
\tablecaption{Expansion Velocities, Asymmetry Factors, and Molecular Production Rates in Wirtanen for a Hemispheric Model\label{tab:vexp}}
\tablewidth{0pt}
\tablehead{
\colhead{Execution} & \colhead{UT Date} & \colhead{\textit{T}\subs{rot}\sups{a}} & \colhead{\textit{v}\subs{exp}(1)\sups{b}} & \colhead{\textit{v}\subs{exp}(2)\sups{c}} & \colhead{\textit{Q}\subs{1}/\textit{Q}\subs{2}\sups{d}} & \colhead{$\chi^2$\subs{red}\sups{e}} & \colhead{\textit{Q}\sups{f}} \\
\colhead{} & \colhead{(2018)}  & \colhead{(K)} & \colhead{(km s\sups{-1})} & \colhead{(km s\sups{-1})} & \colhead{} & \colhead{} & \colhead{(10$^{26}$ s$^{-1}$)}
}
\startdata
1 & 12/7-8, 23:57--00:58  & (82) & 0.76 $\pm$ 0.01 & 0.45 $\pm$ 0.02 & 3.6 $\pm$ 0.2 & 1.35 & 3.6 $\pm$ 0.4 \\
2 & 12/8, 01:26--02:26  & (77) & 0.72 $\pm$ 0.03 & 0.39 $\pm$ 0.03 & 1.5 $\pm$ 0.1 & 1.04 & 2.0 $\pm$ 0.2 \\
\hline
3 & 12/9, 00:46--01:47  & (52) & 0.74 $\pm$ 0.01 & 0.41 $\pm$ 0.02 & 3.4 $\pm$ 0.2 & 1.11 & 2.0 $\pm$ 0.2 \\
4 & 12/9, 02:26--03:27  & (60) & 0.80 $\pm$ 0.01 & 0.45 $\pm$ 0.03 & 5.4 $\pm$ 0.3 & 1.28 & 3.1 $\pm$ 0.3 \\
5 & 12/9, 03:54--04:55  & (57) & 0.73 $\pm$ 0.02 & 0.34 $\pm$ 0.02 & 3.4 $\pm$ 0.3 & 0.97 & 2.0 $\pm$ 0.2
\enddata
\tablecomments{Best-fit parameters for the hemispheric outgassing model. \sups{a} We set the model kinetic temperature to the derived rotational temperature for each execution (see Section~\ref{subsec:trot}). \sups{b} Expansion velocity in the sunward hemisphere. \sups{c} Expansion velocity in the anti-sunward hemisphere. \sups{d} Asymmetry factor: ratio of production rates in the sunward vs.\ anti-sunward hemisphere. \sups{e} Reduced $\chi^2$ of the model fit. \sups{f} CH$_3$OH production rate.
}
\end{deluxetable*}

\section{Variation in Methanol Outgassing and Production Rates}\label{sec:variations}
Table~\ref{tab:vexp} and Figure~\ref{fig:spectra} show that the line velocity profile for CH$_3$OH emission in Wirtanen varied over the course of our observations. As Table~\ref{tab:obslog} and the time stamps in Figure~\ref{fig:spectra} show, Executions 1 and 4, as well as Executions 2 and 5, were separated by $\sim$26.5 hr. The nucleus rotation period of Wirtanen is well constrained from measurements at optical and millimeter wavelengths during the 2018 perihelion passage: \cite{Farnham2018}, \cite{Jehin2018}, \cite{Handzlik2019}, and \cite{Farnham2021} reported \textit{P} = 8.91 $\pm$ 0.03 hr, \textit{P} = 9.19 hr, \textit{P} = 9 hr, and \textit{P} = 9.13 hr, respectively. Coupled with this $\sim$9 hr rotational period, the line morphologies in Figure~\ref{fig:spectra} and the production rates in Table~\ref{tab:vexp} suggest that these effects could be periodic and related to the rotation of Wirtanen's nucleus, as Executions 1 and 4 as well as Executions 2 and 5 would be separated by approximately three complete rotations. These variations may be consistent with changes in coma outgassing as active regions rotated from the day side to the night side of Wirtanen's nucleus. Examining trends in integrated flux (Table~\ref{tab:lines}) and \textit{Q}(CH$_3$OH) (Table~\ref{tab:vexp}) compared to trends in weather conditions (Table~\ref{tab:obslog}) reveals no discernible link between the weather and the observed variability. We examine potential mechanisms for variability further below.

\begin{figure*}[h!]
\plotone{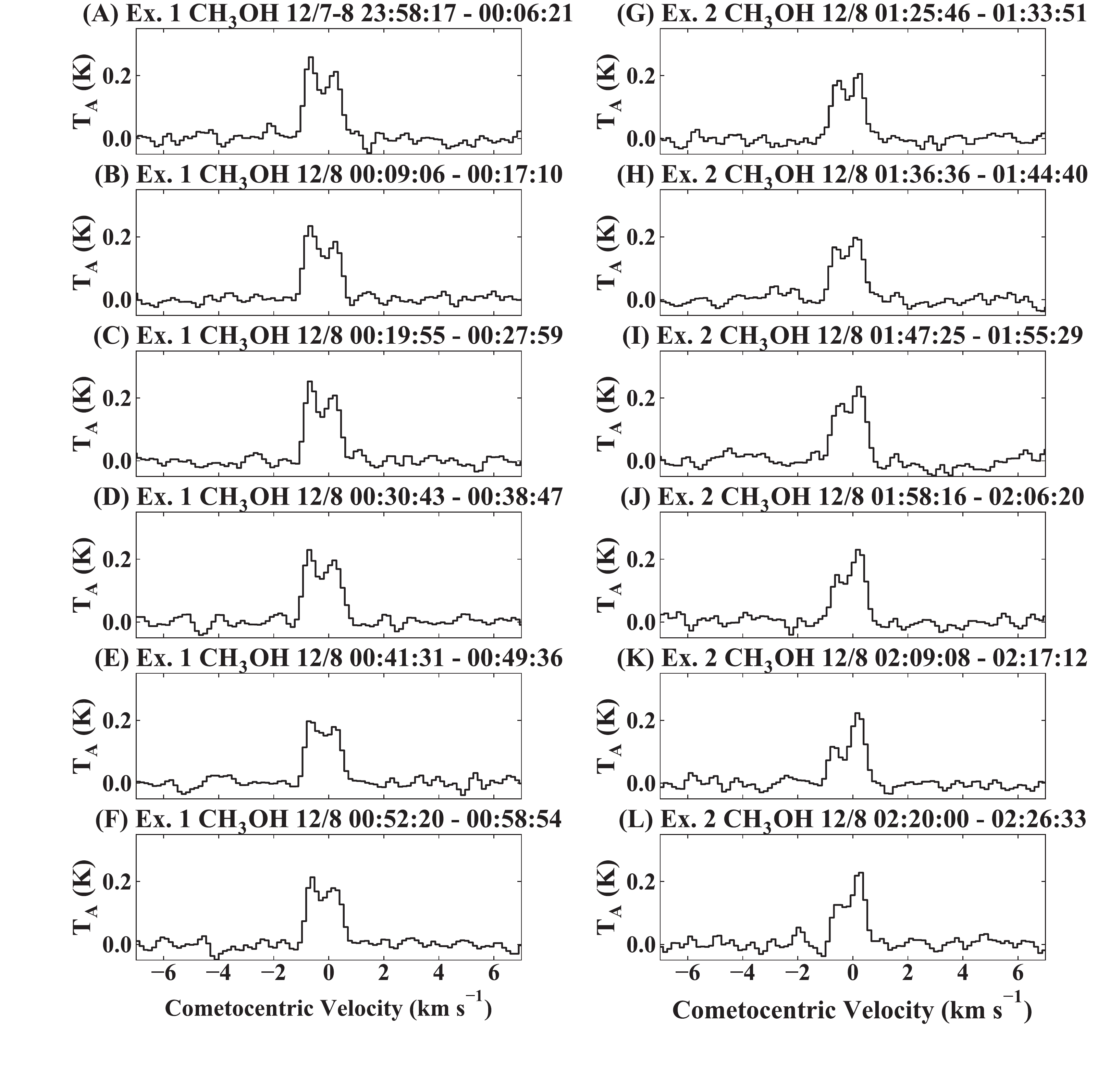}
\caption{Averaged spectra of the CH$_3$OH $5_0$--$4_0$ $E$, $5_{-1}$--$4_{-1}$ $E$, and $5_0$--$4_0$ $A^{+}$ transitions in each scan for Execution 1 (Ex. 1, left column) and Execution 2 (Ex. 2, right column) on December 8 (see Table~\ref{tab:obslog}), showing the evolution in the velocity profile of CH$_3$OH outgassing over each $\sim$8 minute scan.
\label{fig:Dec8subscans}}
\end{figure*}

\subsection{Variation on Timescales of Minutes}\label{subsec:minutes}
\begin{figure*}[h]
\plotone{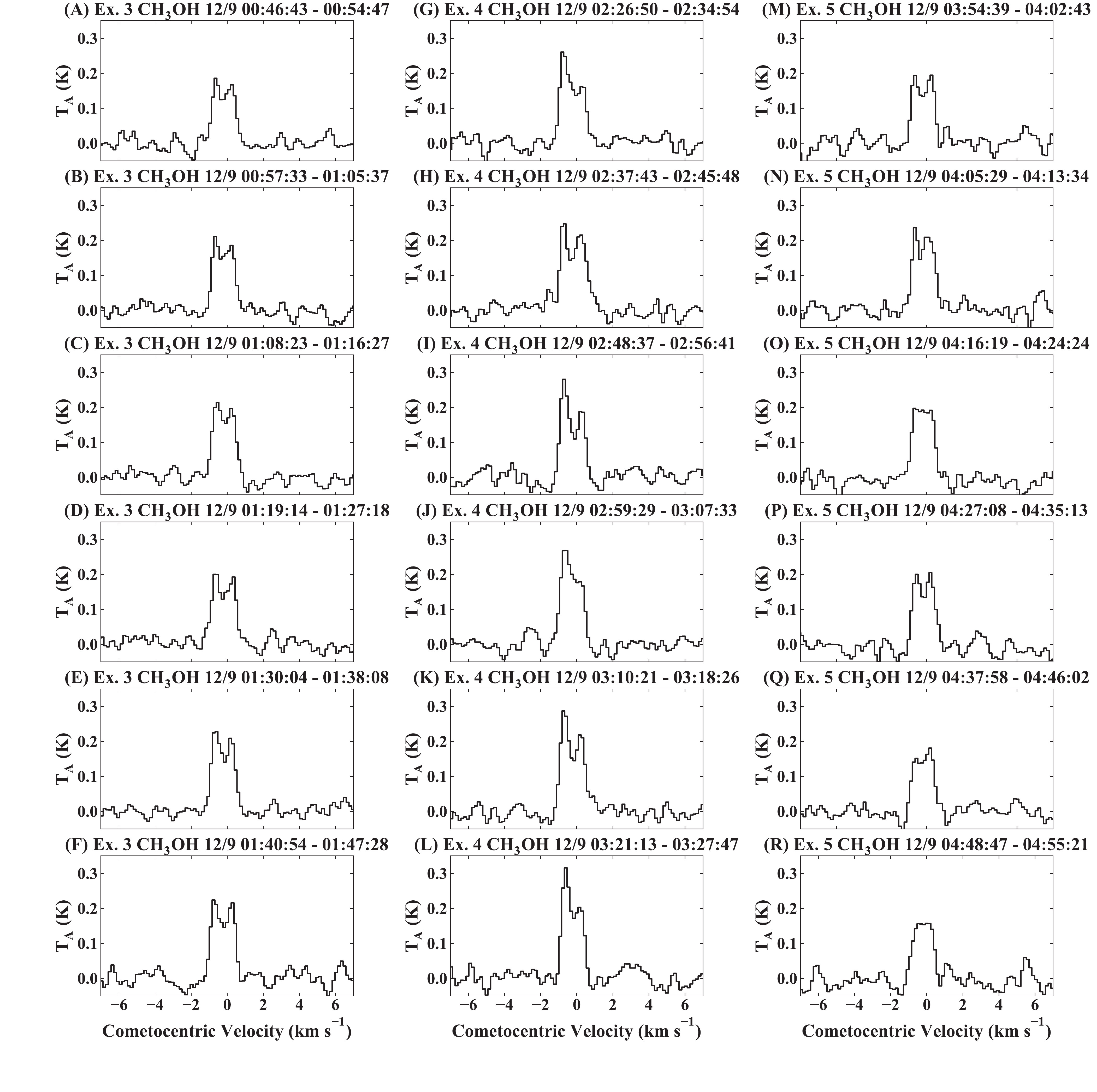}
\caption{Averaged spectra of the CH$_3$OH $5_0$--$4_0$ $E$, $5_{-1}$--$4_{-1}$ $E$, and $5_0$--$4_0$ $A^{+}$ transitions in each scan for Execution 3 (Ex. 3, left column), Execution 4 (Ex. 4, middle column), and Execution 5 (Ex. 5, right column) on December 9 (see Table~\ref{tab:obslog}), showing the evolution in the velocity profile of CH$_3$OH outgassing over each $\sim$8 minute scan. \label{fig:Dec9subscans}}
\end{figure*}

To further address these potential rotational effects, we searched for changes in outgassing on smaller timescales. Each ACA execution consisted of six scans, with each scan lasting $\sim$8 minutes. The S/N in the CH$_3$OH $5_0$--$4_0$ $A^{+}$ transition in individual scans was not high enough to draw definitive conclusions, so we averaged the line profiles of the three strongest transitions in each scan (namely the $5_0$--$4_0$ $E$, $5_{-1}$--$4_{-1}$ $E$, and $5_0$--$4_0$ $A^{+}$ transitions) to improve S/N. Figure~\ref{fig:Dec8subscans} shows our results for December 8, and Figure~\ref{fig:Dec9subscans} shows the same for December 9.

Figure~\ref{fig:Dec8subscans} shows a clear progression from an asymmetric profile with a blueshifted peak (Figure~\ref{fig:spectra}A) to a redshifted peak (Figure~\ref{fig:spectra}B) on December 8. The progression from scan to scan in Figure~\ref{fig:Dec9subscans} on December 9 is less dramatic than on December 8, perhaps owing to the poorer weather (zenith PWV was higher overall on December 9 compared to December 8, and increased over the course of our observations on December 9; see Table~\ref{tab:obslog}). Regardless, the progression from a roughly symmetric profile in Execution 3, to an asymmetric profile with a blueshifted peak and higher amplitude in Execution 4, followed by a decrease in amplitude and return to a more symmetric profile in Execution 5, is clearly visible. 

\begin{figure*}[h]
\plotone{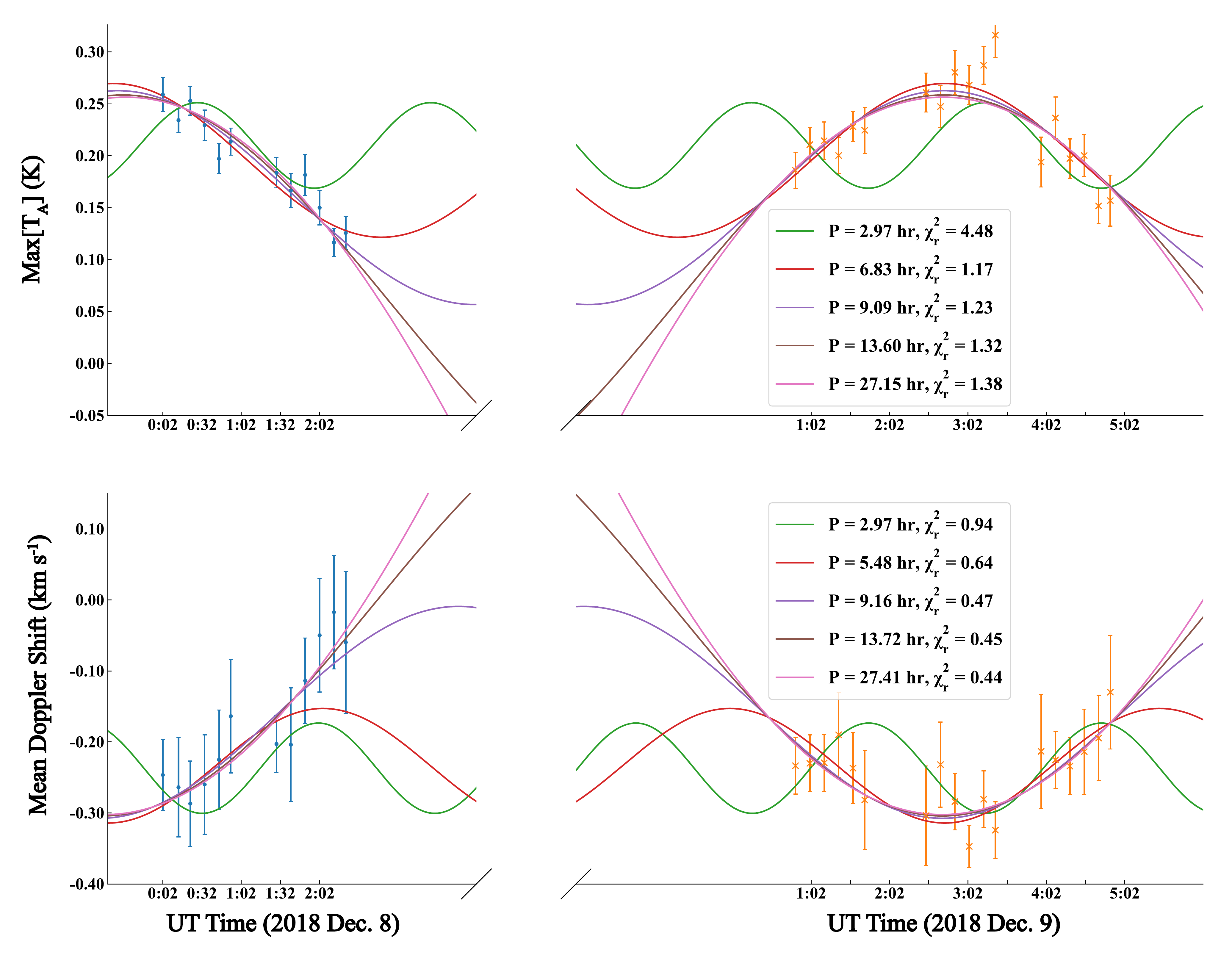}
\caption{Top: maximum amplitude of the blueward peaks (Figure~\ref{fig:Dec8subscans}) in each scan on December 8 and 9 as a function of UT time. Overlaid are best-fit sine waves with periods near 3, 6, 9, 12, and 27 hr, as well as the reduced $\chi^2$ for each fit. Bottom: mean doppler shift (velocity first moment) in each scan on December 8 and 9 as a function of UT time. Overlaid are best-fit sine waves with periods near 3, 6, 9, 12, and 27 hr, as well as the reduced $\chi^2$ for each fit.
\label{fig:phase}}
\end{figure*}

\begin{figure}[h]
\plotone{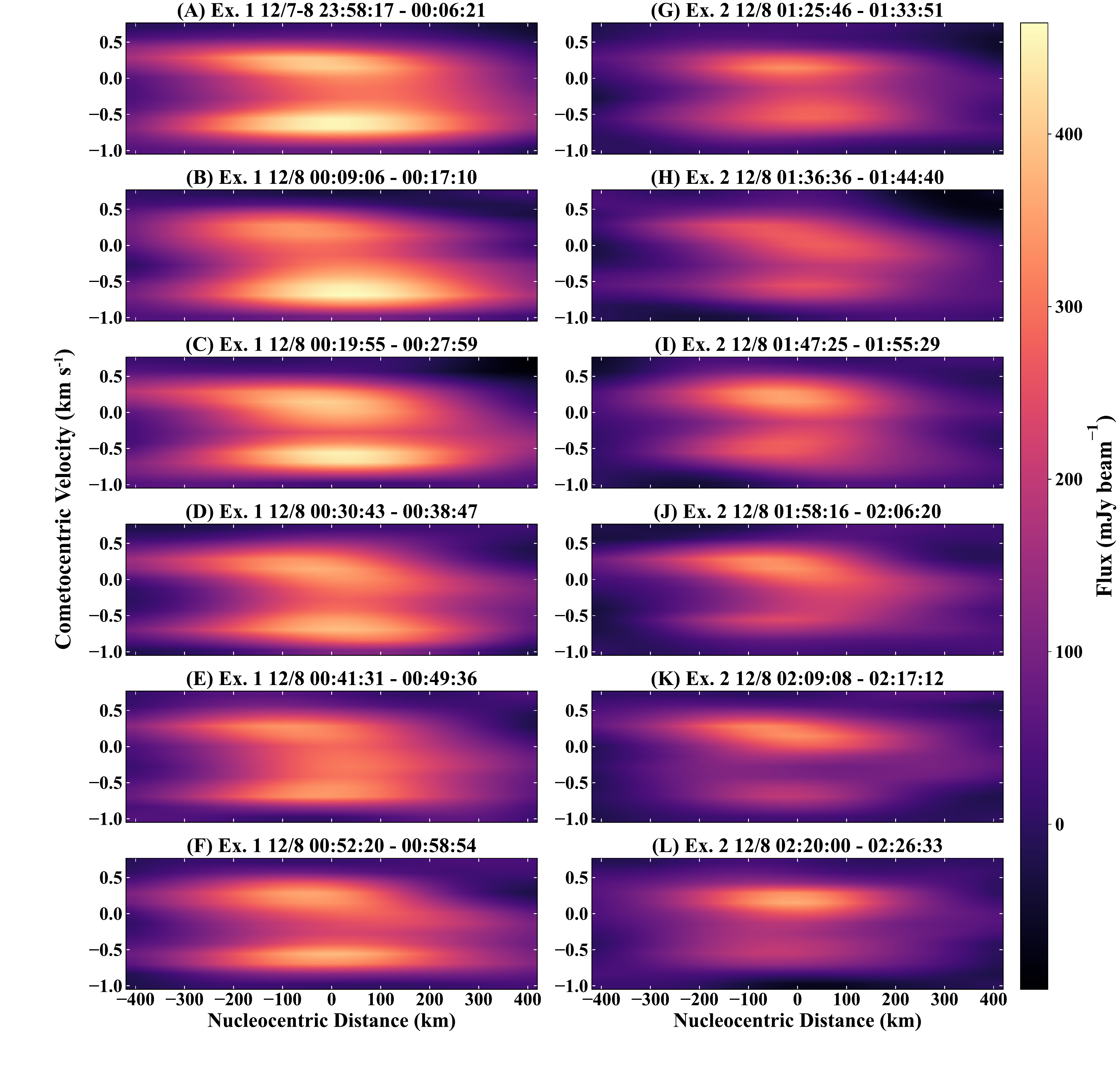}
\caption{Position--velocity diagrams of CH$_3$OH in Wirtanen for each scan on December 8, taken parallel to the solar vector and with the spatial origin at the peak emission position for each scan. Each diagram shows the cometocentric velocity as a function of distance from the image center, with each pixel color-coded by the line flux, thereby showing the evolution in velocity structure across each scan. Scans are labeled and ordered as in Figure~\ref{fig:Dec8subscans}.  Negative offsets are in the anti-sunward direction.
\label{fig:Dec8PV}}
\end{figure}

\begin{figure}[h]
\plotone{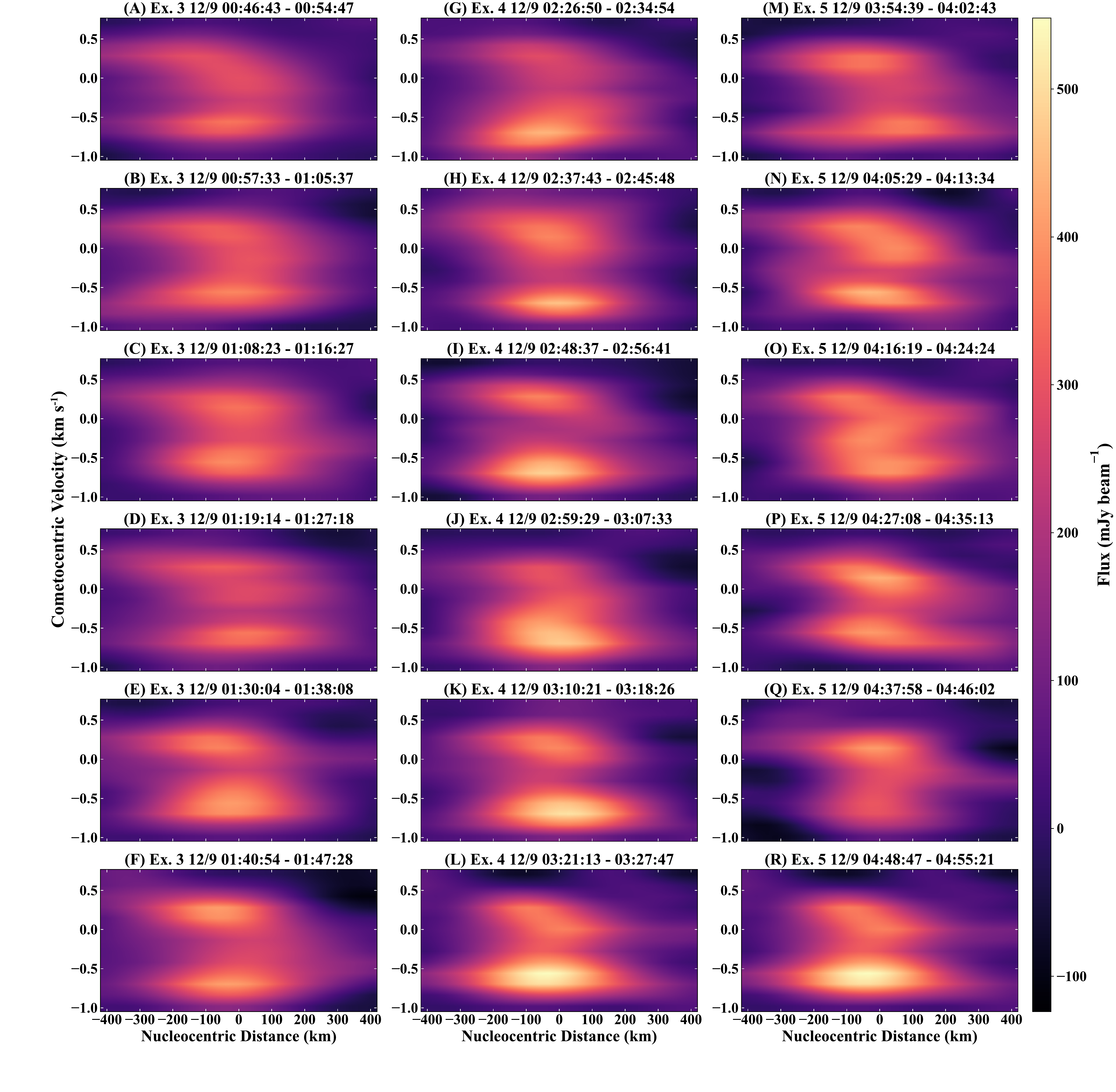}
\caption{Position--velocity diagrams of CH$_3$OH in Wirtanen for each scan on December 9, taken parallel to the solar vector and with the spatial origin at the peak emission position for each scan. Each diagram shows the cometocentric velocity as a function of distance from the image center, with each pixel color-coded by the line flux, thereby showing the evolution in velocity structure across each scan. Scans are labeled and ordered as in Figure~\ref{fig:Dec9subscans}. Negative offsets are in the anti-sunward direction.
\label{fig:Dec9PV}}
\end{figure}

To investigate the potential periodicity of the variation in line asymmetry, we examined the mean Doppler shift (velocity first moment; i.e., the average velocity weighted by the emission intensity in each channel) as well as the maximum amplitude of the blueward peak (taken as the maximum value for \textit{v} $<$ 0 km s$^{-1}$) in each scan on both dates and plotted them as a function of time (Figure~\ref{fig:phase}). We calculated least-squares fits for sine waves with periods near 3, 6, 9, 12, and 27 hr for both plots and calculated the goodness-of-fit (reduced $\chi^2$). The constraint Max[T\subs{A}] (K) $\ge$ 0 excludes \textit{P} = 13.6 hr and 27.15 hr. Based on the reduced $\chi^2$, the amplitude of the blue peaks is best fit by a sine wave with \textit{P} = 6.83 $\pm$ 0.04 hr, amplitude \textit{A} = -0.07 $\pm$ 0.01 K, and \textit{A}$_0$ = 0.20 $\pm$ 0.01 K. Similarly, the mean Doppler shifts are best fit by a sine wave with \textit{P} = 9.16 $\pm$ 0.08 hr, \textit{A} = -0.15 $\pm$ 0.02 km s$^{-1}$, and \textit{v}$_0$ = -0.16 $\pm$ 0.01 km s$^{-1}$ (when excluding \textit{P} = 13 or 27 hr). Thus, within the limitations of the data, our results suggest \textit{P} $\sim$ 7--9 hr, consistent with measurements from optical and millimeter wavelengths \citep[\textit{P} $\sim$ 9 hr;][]{Farnham2018,Jehin2018,Handzlik2019,Farnham2021}. The consistent periodicity between our observations and those from optical and millimeter wavelengths suggests that the variation in CH$_3$OH outgassing in Wirtanen reported here is indeed tied to rotational effects. The rotational variation in the velocity structure in Wirtanen is also clearly evident in the position-velocity diagrams for each scan (Figures~\ref{fig:Dec8PV} and ~\ref{fig:Dec9PV}).

In addition to its measured period, the nucleus of Wirtanen is well studied due to its status as the original target for the Rosetta mission. It is a small (\textit{r}\subs{eff} $\sim$ 0.5 km) prolate ellipsoid \citep[axial ratio $>$ 1.4;][]{Boehnhardt2002}. Although previous observations were unable to discern whether Wirtanen is in a state of non-principal axis rotation \citep{Samarasinha1996}, measurements during the 2018 perihelion passage demonstrated that Wirtanen is in a state of simple rotation \citep{Farnham2021}. Given Wirtanen's prolate elliptical nucleus shape and rotational period, coupled with the periodicity measured in our observations, it is possible that the variations in CH$_3$OH outgassing are due to differences in illumination on the nucleus, along with active sites rotating from the day side to the night side. Such rotational effects have been observed in a comet with a prolate elliptical nucleus before, namely 103P/Hartley 2. At 103P, the EPOXI mission, coupled with ground-based observations with IRAM and CSO, found that varying illumination of the small lobe of the nucleus due to rotation (along with the nucleus shape) caused significant changes in volatile release \citep{AHearn2011,Drahus2012,Boissier2014}. Furthermore, at comet 67P/Churyumov-Gerasimenko, the Rosetta mission found diurnal variations in CH$_4$, CO, and C$_2$H$_6$ outgassing \citep{Luspay-Kuti2015,Fink2016} tied to the rotation of the comet's bilobed nucleus. Finally, in comet C/1995 O1 (Hale-Bopp), \cite{Bockelee2009} reported strong variations in the spectral line profile of CO consistent with a rotating jet. This rotating jet spiraled along with the rotating nucleus and was active during both the day and night.  Clearly, nucleus rotation and complex outgassing are strongly linked in multiple comets.

\subsection{Asymmetric Line Center and Hyperactivity}\label{subsec:linecenter}
In light of the rapidly varying asymmetry in Wirtanen's CH$_3$OH spectral line profile, an equally striking feature is the consistent blueward offset of the line center observed in \textit{every} spectrum across the two dates of our measurements, including in individual 8 minute scans. Coupled with the higher expansion velocity in the sunward hemisphere compared to the anti-sunward hemisphere (Table~\ref{tab:vexp}), this suggests that there is an excess of CH$_3$OH outflow on the sunward side of the nucleus. Furthermore, the redward side of the profile is remarkably stable across all of our observations, even including Execution 2 on December 8 (Figure~\ref{fig:Dec8subscans}), where it retains a relatively constant amplitude as the blueward peak continually diminishes. This implies that the overall variability in CH$_3$OH production may be largely due to activity occurring on the dayward side. This is consistent with optical observations of CN in Wirtanen that indicated two sources of CN production: one which remained active throughout a rotation, and a second which turned on and off as the nucleus rotated \citep{Farnham2021}.

Wirtanen is considered a hyperactive comet \citep[e.g.,][and references therein]{Lis2019}, with an unusually high \textit{Q}(H$_2$O) for its small nucleus size attributed to a significant release of H$_2$O from icy grains in the coma. Observations of H$_2$O hot bands near 2.9 $\mu$m with NIRSPEC-2 at the W. M. Keck Observatory \citep{Martin2016,Martin2018} on December 17 and 18 \citep{Bonev2021} and iSHELL at the NASA-IRTF \citep{Rayner2012,Rayner2016} on December 6, December 18 \citep{Roth2021}, and December 21 \citep{Khan2021} found significant enhancements in the spatial profiles of H$_2$O emission, supporting the icy grain contribution to Wirtanen's H$_2$O production \citep[as much as 40\% from extended sources;][]{Bonev2021}. 

\begin{figure}[h]
\plotone{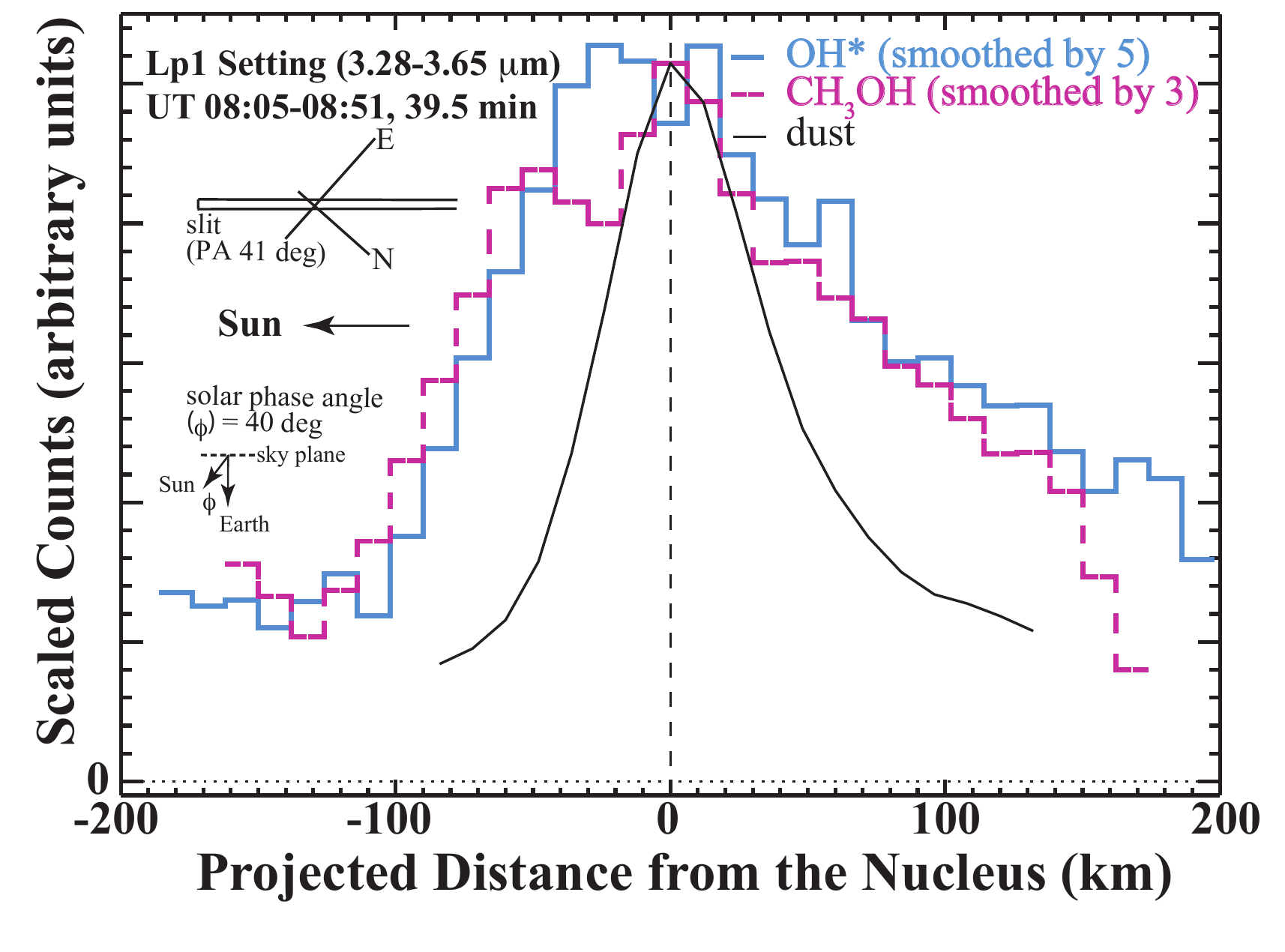}
\caption{Spatial profiles of emission for CH$_3$OH, H$_2$O (OH*), and dust continuum in 46P/Wirtanen measured with iSHELL on UT 2018 December 6. The slit was oriented along the Sun--comet line (PA 41$\degr$), with the direction of the Sun indicated. Also shown is the Sun--comet--Earth angle (phase angle, $\phi$ = 39$\degr$). From observations obtained through NASA-IRTF Director's Discretionary Time, in support of the 46P ALMA campaign.
\label{fig:iSHELL}}
\end{figure}

Given its hyperactivity, an intriguing possible explanation for the consistent blueward offset in the CH$_3$OH line center is enhanced sublimation on the sunward side of the nucleus, perhaps from icy grains in the coma. Although the presence of icy grains would change the absolute \textit{Q}(CH$_3$OH) for each execution (compared to the parent species models used here), the trends in production rate and line morphology would remain the same. This is consistent with measurements of Wirtanen at other wavelengths during its 2018 perihelion passage. \cite{Cordiner2019b} reported extended CH$_3$OH release in Wirtanen on UT 2018 December 7, most likely from an icy grain source, based on ALMA observations with the main 12 m array. \cite{Wang2020} found that the HCN (\textit{J} = 1--0) transition near 3.4 mm displayed a strong blueward offset in its line center on December 14 and 15, consistent with anisotropic outgassing on its sunward side. Furthermore, simultaneous iSHELL measurements of CH$_3$OH, OH* \citep[prompt emission, a direct tracer of H$_2$O production; see][]{Bonev2006}, and dust continuum taken on December 6 found that CH$_3$OH and H$_2$O displayed significant peaks in emission in the sunward-facing hemisphere (Figure~\ref{fig:iSHELL}). Post-perihelion measurements with NIRSPEC-2 \citep{Bonev2021} and iSHELL \citep{Roth2021,Khan2021} showed that the enhancements in CH$_3$OH and H$_2$O persisted, although now in the projected anti-sunward direction (but at a considerably smaller phase angle, $\phi \sim 18\degr$). If the CH$_3$OH enhancement in Wirtanen is indeed owing to icy grain sublimation, it is not the first time that icy grain contributions to multiple species have been found in a hyperactive comet. In comet 103P/Hartley 2, \cite{Drahus2012} found that HCN displayed a varying blueshifted peak corresponding to production from a jet, while the remaining HCN production was attributed to icy grains. In contrast, CH$_3$OH did not display a blueshifted peak, suggesting the CH$_3$OH was produced solely from icy grains. Measurements by \cite{Boissier2014} also indicated the production of HCN and CH$_3$OH from icy grains in 103P/Hartley 2, implied by the phase offset between the CO$_2$ production curve (produced purely from nucleus sources) and the HCN, CH$_3$OH, and H$_2$O production curves. The similarities between these two comets suggest that their outgassing behaviors and sources may have been similar, and emphasize the complex comae of hyperactive comets. Future work analyzing our full ALMA data set of observations toward Wirtanen will enable us to clarify the extent to which icy grain sublimation contributed to CH$_3$OH production in its coma.

\section{Conclusion}\label{sec:conclusion}
The historic 2018 perihelion passage of Wirtanen together with the capabilities of ALMA enabled us to measure its inner coma with high spatial, spectral, and temporal resolution. Our ALMA ACA observations presented here revealed rapidly varying and anisotropic CH$_3$OH production in the inner coma of Wirtanen on two pre-perihelion dates shortly before its closest approach to Earth, with the evolution of the spectral line asymmetry clearly visible on a timescale of minutes. We found that the variations in CH$_3$OH were periodic, in good agreement with the nucleus rotational period derived from optical and millimeter wavelength measurements (\textit{P} $\sim$ 9 hr), and consistent with hemispheric outgassing along the Sun--comet vector, suggesting variations in outgassing owing to changing illumination of active sites on the nucleus. Furthermore, the consistent blueward offset of the CH$_3$OH line center, coupled with the higher expansion velocity in the sunward hemisphere, suggested enhanced CH$_3$OH sublimation in this direction, perhaps from icy grains. The work presented here demonstrates that time-resolved studies of the composition, spatial extent, and outgassing profiles of coma species are possible with the high sensitivity of ALMA on timescales as small as minutes, and lays the groundwork for future detailed cometary coma studies using ALMA ACA.

\acknowledgments
This work was supported by the NASA Postdoctoral Program at the NASA Goddard Space Flight Center, administered by Universities Space Research Association under contract with NASA (NXR), as well as the National Science Foundation (under grant No. AST-1614471; MAC), and by the Planetary Science Division Internal Scientist Funding Program through the Fundamental Laboratory Research (FLaRe) work package (SNM, MAC, SBC), as well as the NASA Astrobiology Institute through the Goddard Center for Astrobiology (proposal 13-13NAI7-0032; SNM, MAC, SBC). The National Science Foundation supported this work (grants \#1616306 and \#2009398; BPB). EJ is a F.R.S.-FNRS Belgian Senior Research Associate. Funding is gratefully acknowledged through the NASA Planetary Astronomy/Solar System Observations Program (grant \#18-SSO18\_2-0040, M.A.D.; grant 80NSSC17K0705, B.P.B. and N.D.R.) This study makes use of the following ALMA data: ADS/JAO.ALMA \#2018.1.01114.S. ALMA is a partnership of ESO (representing its member states), NSF (USA), and NINS (Japan), together with NRC (Canada), MOST and ASIAA (Taiwan), and KASI (Republic of Korea) in cooperation with the Republic of Chile. The Joint ALMA Observatory is operated by ESO, AUI/NRAO, and NAOJ. The National Radio Astronomy Observatory is a facility of the National Science Foundation operated under cooperative agreement by Associated Universities, Inc. Data for this study were obtained at the NASA Infrared Telescope Facility (IRTF), operated by the University of Hawaii under contract NNH14CK55B with the National Aeronautics and Space Administration. We are most fortunate to the have the opportunity to conduct observations from Maunakea and recognize the very significant cultural role and reverence that the summit of Maunakea has always had within the indigenous community. We thank anonymous reviewers for suggestions that improved this manuscript.

\pagebreak
\bibliography{46P}{}

\begin{thebibliography}{}
\expandafter\ifx\csname natexlab\endcsname\relax\def\natexlab#1{#1}\fi
\providecommand{\url}[1]{\href{#1}{#1}}
\providecommand{\dodoi}[1]{doi:~\href{http://doi.org/#1}{\nolinkurl{#1}}}
\providecommand{\doeprint}[1]{\href{http://ascl.net/#1}{\nolinkurl{http://ascl.net/#1}}}
\providecommand{\doarXiv}[1]{\href{https://arxiv.org/abs/#1}{\nolinkurl{https://arxiv.org/abs/#1}}}

\bibitem[{{A'Hearn} {et~al.}(1995){A'Hearn}, {Millis}, {Schleicher}, {Osip}, \&
  {Birch}}]{AHearn1995}
{A'Hearn}, M.~F., {Millis}, R.~L., {Schleicher}, D.~G., {Osip}, D.~J., \&
  {Birch}, V.~P. 1995, Icarus, 118, 223

\bibitem[{A'Hearn {et~al.}(2011)A'Hearn, Belton, Delamere, Feaga, Hampton,
  Kissel, Klaase, McFadden, Meech, Melosh, Schultz, Sunshine, Thomas, Veverka,
  Wellnitz, Yeomans, Besse, Bodewits, Bowling, Carcich, Collins, Farnham,
  Groussin, Hermalyn, Kelley, Kelley, Li, Lindler, Lisse, McLaughlin, Merlin,
  Protopapa, Richardson, \& Williams}]{AHearn2011}
A'Hearn, M.~F., Belton, M. J.~S., Delamere, W.~A., {et~al.} 2011, Science, 332,
  1396

\bibitem[{{Biver} {et~al.}(1999){Biver}, {Bockel{\'e}e-Morvan}, {Crovisier},
  {Davies}, {Matthews}, {Wink}, {Rauer}, {Colom}, {Dent}, {Despois}, {Moreno},
  {Paubert}, {Jewitt}, \& {Senay}}]{Biver1999}
{Biver}, N., {Bockel{\'e}e-Morvan}, D., {Crovisier}, J., {et~al.} 1999, AJ,
  118, 1850

\bibitem[{{Biver} {et~al.}(2021){Biver}, {Bockel{\'e}e-Morvan}, {Boissier},
  {Moreno}, {Crovisier}, {Lis}, {Colom}, {Cordiner}, {Milam}, {Roth}, {Bonev},
  {Dello Russo}, {Vervack}, \& {DiSanti}}]{Biver2021}
{Biver}, N., {Bockel{\'e}e-Morvan}, D., {Boissier}, J., {et~al.} 2021, A\&A, In
  Press

\bibitem[{{Bockel{\'e}e-Morvan} \& {Biver}(2017)}]{Bockelee2017}
{Bockel{\'e}e-Morvan}, D., \& {Biver}, N. 2017, PTRSA, 375, 20160252

\bibitem[{{Bockel{\'e}e-Morvan} {et~al.}(1994){Bockel{\'e}e-Morvan},
  {Crovisier}, {Colom}, \& {Despois}}]{Bockelee1994}
{Bockel{\'e}e-Morvan}, D., {Crovisier}, J., {Colom}, P., \& {Despois}, D. 1994,
  A\&A, 287, 647

\bibitem[{{Bockel{\'e}e-Morvan} {et~al.}(2004){Bockel{\'e}e-Morvan},
  {Crovisier}, {Mumma}, \& {Weaver}}]{Bockelee2004}
{Bockel{\'e}e-Morvan}, D., {Crovisier}, J., {Mumma}, M.~J., \& {Weaver}, H.~A.
  2004, in Comets II, ed. H.~U. {Keller} \& H.~A. {Weaver} (University of
  Arizona Press), 391

\bibitem[{{Bockel{\'e}e-Morvan} {et~al.}(2009){Bockel{\'e}e-Morvan}, {Henry},
  {Biver}, {Boissier}, {Colom}, {Crovisier}, {Despois}, {Moreno}, \&
  {Wink}}]{Bockelee2009}
{Bockel{\'e}e-Morvan}, D., {Henry}, F., {Biver}, N., {et~al.} 2009, A\&A, 505,
  825

\bibitem[{{Boehnhardt} {et~al.}(2002){Boehnhardt}, {Delahodde}, {Sekiguchi},
  {Tozzi}, {Amestica}, {Hainaut}, {Spyromilio}, {Tarenghi}, {West}, {Schulz},
  \& {Schwehm}}]{Boehnhardt2002}
{Boehnhardt}, H., {Delahodde}, C., {Sekiguchi}, T., {et~al.} 2002, A\&A, 387,
  1107

\bibitem[{{Boissier} {et~al.}(2007){Boissier}, {Bockel{\'e}e-Morvan}, {Biver},
  {Crovisier}, {Despois}, {Marsden}, \& {Moreno}}]{Boissier2007}
{Boissier}, J., {Bockel{\'e}e-Morvan}, D., {Biver}, N., {et~al.} 2007, A\&A,
  475, 1131

\bibitem[{Boissier {et~al.}(2014)Boissier, Bockel{\'e}e-Morvan, Biver, Colom,
  Crovisier, Moreno, Zakharov, Groussin, Jorda, \& Lis}]{Boissier2014}
Boissier, J., Bockel{\'e}e-Morvan, D., Biver, N., {et~al.} 2014, Icarus, 228,
  197

\bibitem[{Bonev {et~al.}(2006)Bonev, Mumma, DiSanti, Dello~Russo, Magee-Sauer,
  Ellis, \& Stark}]{Bonev2006}
Bonev, B.~P., Mumma, M.~J., DiSanti, M.~A., {et~al.} 2006, ApJ, 653, 774

\bibitem[{{Bonev} {et~al.}(2021){Bonev}, {Dello Russo}, {DiSanti}, {Martin},
  {Doppmann}, {Vervack}, {Villanueva}, {Kawakita}, {Gibb}, {Combi}, {Roth},
  {Saki}, {McKay}, {Cordiner}, {Bodewits}, {Crovisier}, {Biver}, {Cochran},
  {Shou}, {Khan}, \& {Venkataramani}}]{Bonev2021}
{Bonev}, B.~P., {Dello Russo}, N., {DiSanti}, M.~A., {et~al.} 2021, PSJ, 2, 45

\bibitem[{{Brinch} \& {Hogerheijde}(2010)}]{Brinch2010}
{Brinch}, C., \& {Hogerheijde}, M.~R. 2010, A\&A, 523, A25

\bibitem[{Cochran {et~al.}(2015)Cochran, Levasseur-Regourd, Cordiner, Hadamcik,
  Lasue, Gicquel, Schleicher, Charnley, Mumma, Paganini, Bockel{\'e}e-Morvan,
  Biver, \& Kuan}]{Cochran2015}
Cochran, A.~L., Levasseur-Regourd, A.-C., Cordiner, M., {et~al.} 2015, SSRv,
  197, 9

\bibitem[{{Cordiner} {et~al.}(2014){Cordiner}, {Remijan}, {Boissier}, {Milam},
  {Mumma}, {Charnley}, {Paganini}, {Villanueva}, {Bockel{\'e}e-Morvan}, {Kuan},
  {Chuang}, {Lis}, {Biver}, {Crovisier}, {Minniti}, \&
  {Coulson}}]{Cordiner2014}
{Cordiner}, M.~A., {Remijan}, A.~J., {Boissier}, J., {et~al.} 2014, ApJL, 792,
  L2

\bibitem[{{Cordiner} {et~al.}(2017a){Cordiner}, {Biver}, {Crovisier},
  {Bockel{\'e}e-Morvan}, {Mumma}, {Charnley}, {Villanueva}, {Paganini}, {Lis},
  {Milam}, {Remijan}, {Coulson}, {Kuan}, \& {Boissier}}]{Cordiner2017a}
{Cordiner}, M.~A., {Biver}, N., {Crovisier}, J., {et~al.} 2017a, ApJ, 837, 177

\bibitem[{{Cordiner} {et~al.}(2017b){Cordiner}, {Boissier}, {Charnley},
  {Remijan}, {Mumma}, {Villanueva}, {Lis}, {Milam}, {Paganini}, {Crovisier},
  {Bockel{\'e}e-Morvan}, {Kuan}, {Biver}, \& {Coulson}}]{Cordiner2017b}
{Cordiner}, M.~A., {Boissier}, J., {Charnley}, S.~B., {et~al.} 2017b, ApJ, 838,
  147

\bibitem[{{Cordiner} {et~al.}(2019{\natexlab{a}}){Cordiner}, {Palmer}, {de
  Val-Borro}, {Charnley}, {Paganini}, {Villanueva}, {Bockel{\'e}e-Morvan},
  {Biver}, {Remijan}, {Kuan}, {Milam}, {Crovisier}, {Lis}, \&
  {Mumma}}]{Cordiner2019}
{Cordiner}, M.~A., {Palmer}, M.~Y., {de Val-Borro}, M., {et~al.}
  2019{\natexlab{a}}, ApJL, 870, L26

\bibitem[{{Cordiner} {et~al.}(2019{\natexlab{b}}){Cordiner}, {Biver}, {Milam},
  {Charnley}, {Remijan}, {de Val-Borro}, {Bonev}, {Mumma}, {Villanueva},
  {Paganini}, {Bockel{\'e}e-Morvan}, {Crovisier}, {Kuan}, {Lis}, {Boissier}, \&
  {Qi}}]{Cordiner2019b}
{Cordiner}, M.~A., {Biver}, N., {Milam}, S.~N., {et~al.} 2019{\natexlab{b}}, in
  EPS-DPS Joint Meeting 2019, Vol.~13, EPSC Abstracts, EPSC, Geneva,
  Switzerland, EPSC--DPS2019--1131--2

\bibitem[{{Crovisier} {et~al.}(2009){Crovisier}, {Biver},
  {Bockel{\'e}e-Morvan}, \& {Colom}}]{Crovisier2009}
{Crovisier}, J., {Biver}, N., {Bockel{\'e}e-Morvan}, D., \& {Colom}, P. 2009,
  P\&SS, 57, 1162

\bibitem[{Dello~Russo {et~al.}(2016a)Dello~Russo, Kawakita, Jr., \&
  Weaver~H.}]{DelloRusso2016a}
Dello~Russo, N., Kawakita, H., Jr., V. R.~J., \& Weaver~H., A. 2016a, Icarus,
  278, 301

\bibitem[{Drahus {et~al.}(2012)Drahus, Jewitt, Guilbert-Lepoutre, Waniak, \&
  Sievers}]{Drahus2012}
Drahus, M., Jewitt, D., Guilbert-Lepoutre, A., Waniak, W., \& Sievers, A. 2012,
  ApJ, 756, 80

\bibitem[{{Farnham} {et~al.}(2018){Farnham}, {Knight}, \&
  {Schleicher}}]{Farnham2018}
{Farnham}, T.~L., {Knight}, M.~M., \& {Schleicher}, D.~G. 2018, Central Bureau
  for Astronomical Telegrams, CBET 4571

\bibitem[{{Farnham} {et~al.}(2021){Farnham}, {Knight}, {Schleicher}, {Feaga},
  {Bodewits}, {Skiff}, \& {Schindler}}]{Farnham2021}
{Farnham}, T.~L., {Knight}, M.~M., {Schleicher}, D.~G., {et~al.} 2021, PSJ, 2,
  7

\bibitem[{Fink {et~al.}(2016)Fink, Doose, Rinaldi, Bieler, Capaccioni,
  Bockel{\'e}e-Morvan, Filacchione, Erard, Leyrat, Blecka, Capria, Combi,
  Crovisier, De~Sanctis, Fougere, Taylor, Migliorini, \& Piccioni}]{Fink2016}
Fink, U., Doose, L., Rinaldi, G., {et~al.} 2016, Icarus, 277, 78

\bibitem[{Fray {et~al.}(2006)Fray, B{\'e}nilan, Biver, Bockel{\'e}e-Morvan,
  Cottin, Crovisier, \& Gazeau}]{Fray2006}
Fray, N., B{\'e}nilan, Y., Biver, N., {et~al.} 2006, Icarus, 184, 239

\bibitem[{{Handzlik} {et~al.}(2019){Handzlik}, {Drahus}, \&
  {Kurowski}}]{Handzlik2019}
{Handzlik}, B., {Drahus}, M., \& {Kurowski}, S. 2019, in EPSC-DPS Joint Meeting
  2019, Vol.~13, EPSC Abstracts, EPSC, Geneva, Switzerland,
  EPSC--DPS2019--1775--1

\bibitem[{{Haser}(1957)}]{Haser1957}
{Haser}, L. 1957, BSRSL, 43, 740

\bibitem[{{Huebner} {et~al.}(1992){Huebner}, {Keady}, \& {Lyon}}]{Huebner1992}
{Huebner}, W.~F., {Keady}, J.~J., \& {Lyon}, S.~P. 1992, Ap\&SS, 195, 1

\bibitem[{{Jehin} {et~al.}(2018){Jehin}, {Moulane}, {Manfroid}, \&
  {Pozuelos}}]{Jehin2018}
{Jehin}, E., {Moulane}, Y., {Manfroid}, J., \& {Pozuelos}, F. 2018, Central
  Bureau for Astronomical Telegrams, CBET 4585

\bibitem[{{Khan} {et~al.}(2021){Khan}, {Gibb}, {Bonev}, {Roth}, {Saki},
  {DiSanti}, {Dello Russo}, {Vervack}, {McKay}, \& {Kawakita}}]{Khan2021}
{Khan}, Y., {Gibb}, E.~L., {Bonev}, B.~P., {et~al.} 2021, PSJ, 2, 20

\bibitem[{{Lis} {et~al.}(2008){Lis}, {Bockel{\'e}e-Morvan}, {Boissier},
  {Crovisier}, {Biver}, \& {Charnley}}]{Lis2008}
{Lis}, D.~C., {Bockel{\'e}e-Morvan}, D., {Boissier}, J., {et~al.} 2008, ApJ,
  675, 931

\bibitem[{{Lis} {et~al.}(2019){Lis}, {Bockel{\'e}e-Morvan}, {G{\"u}sten},
  {Biver}, {Stutzki}, {Delorme}, {Dur{\'a}n}, {Wiesemeyer}, \&
  {Okada}}]{Lis2019}
{Lis}, D.~C., {Bockel{\'e}e-Morvan}, D., {G{\"u}sten}, R., {et~al.} 2019, A\&A,
  625, L5

\bibitem[{Luspay-Kuti {et~al.}(2015)Luspay-Kuti, H{\"a}ssig, Fuselier, Mandt,
  Altwegg, Balsiger, Gasc, J{\"a}ckel, Le~Roy, Rubin, Tzou, Wurz, Mousis,
  Dhooghe, Berthelier, Fiethe, Gombosi, \& Mall}]{Luspay-Kuti2015}
Luspay-Kuti, A., H{\"a}ssig, M., Fuselier, S.~A., {et~al.} 2015, A\&A, 583, A4

\bibitem[{{Martin} {et~al.}(2016){Martin}, {Fitzgerald}, {McLean}, {Kress}, \&
  {Wang}}]{Martin2016}
{Martin}, E.~C., {Fitzgerald}, M.~P., {McLean}, I.~S., {Kress}, E., \& {Wang},
  E. 2016, Proc. SPIE, 9908, 99082R

\bibitem[{{Martin} {et~al.}(2018){Martin}, {Fitzgerald}, {McLean}, {Doppmann},
  {Kassis}, {Aliado}, {Canfield}, {Johnson}, {Kress}, {Lanclos}, {Magnone},
  {Sohn}, {Wang}, \& {Weiss}}]{Martin2018}
{Martin}, E.~C., {Fitzgerald}, M.~P., {McLean}, I.~S., {et~al.} 2018, Proc.
  SPIE, 10702, 107020A

\bibitem[{{McMullin} {et~al.}(2007){McMullin}, {Waters}, {Schiebel}, {Young},
  \& {Golap}}]{McMullin2007}
{McMullin}, J.~P., {Waters}, B., {Schiebel}, D., {Young}, W., \& {Golap}, K.
  2007, in Astronomical Data Analysis Software and Systems XVI ASP Conference
  Series, ed. R.~A. {Shaw}, F.~{Hill}, \& D.~J. {Bell}, Vol. 376 (San
  Francisco, CA: ASP), 127

\bibitem[{{Milam} {et~al.}(2006){Milam}, {Remijan}, {Womack}, {Abrell},
  {Ziurys}, {Wyckoff}, {Apponi}, {Friedel}, {Snyder}, {Veal}, {Palmer},
  {Woodney}, {A'Hearn}, {Forster}, {Wright}, {de Pater}, {Choi}, \&
  {Gesmundo}}]{Milam2006}
{Milam}, S.~N., {Remijan}, A.~J., {Womack}, M., {et~al.} 2006, ApJ, 649, 1169

\bibitem[{Mumma \& Charnley(2011)}]{Mumma2011a}
Mumma, M.~J., \& Charnley, S.~B. 2011, ARA\&A, 49, 471

\bibitem[{Rayner {et~al.}(2012)Rayner, Bond, Bonnet, Jaffe, Muller, \&
  Tokunaga}]{Rayner2012}
Rayner, J., Bond, T., Bonnet, M., {et~al.} 2012, Proceedings of the SPIE, 8446,
  84462C

\bibitem[{Rayner {et~al.}(2016)Rayner, Tokunaga, Jaffe, Bonnet, Ching,
  Connelley, Kokubun, Lockhart, \& Warmbier}]{Rayner2016}
Rayner, J., Tokunaga, A., Jaffe, D., {et~al.} 2016, Proceedings of the SPIE,
  9908, 990884

\bibitem[{{Roth} {et~al.}(2021){Roth}, {Bonev}, {DiSanti}, {Dello Russo},
  {McKay}, {Gibb}, {Saki}, {Khan}, {Vervack}, {Kawakita}, {Cochran}, {Biver},
  {Cordiner}, {Crovisier}, {Jehin}, \& {Weaver}}]{Roth2021}
{Roth}, N.~X., {Bonev}, B.~P., {DiSanti}, M.~A., {et~al.} 2021, PSJ, In Press

\bibitem[{{Samarasinha} {et~al.}(1996){Samarasinha}, {Mueller}, \&
  {Belton}}]{Samarasinha1996}
{Samarasinha}, N.~H., {Mueller}, B. E.~A., \& {Belton}, M. J.~S. 1996, P\&SS,
  44, 275

\bibitem[{{Sch{\"o}ier} {et~al.}(2005){Sch{\"o}ier}, {van der Tak}, {van
  Dishoeck}, \& {Black}}]{Schoier2005}
{Sch{\"o}ier}, F.~L., {van der Tak}, F.~F.~S., {van Dishoeck}, E.~F., \&
  {Black}, J.~H. 2005, \aap, 432, 369, \dodoi{10.1051/0004-6361:20041729}

\bibitem[{{Wang} {et~al.}(2020){Wang}, {Zhang}, {Tseng}, {Sun}, {Liao}, {Ip},
  {Zheng}, {Wang}, {Lu}, {Chen}, {Shan}, {Yuan}, {Yan}, \& {Ping}}]{Wang2020}
{Wang}, Z., {Zhang}, S.~B., {Tseng}, W.~L., {et~al.} 2020, AJ, 159, 240

\end{thebibliography}
\bibliographystyle{aasjournal}



\end{document}